\documentclass{article}

\usepackage{microtype}
\usepackage{graphicx}
\usepackage{subfigure}
\usepackage{booktabs} % for professional tables
\usepackage{CJKutf8}

\usepackage{hyperref}
\usepackage{enumitem}

\usepackage[accepted]{icml2021_custom}
\usepackage{icml2021_custom}

\setcitestyle{square}
\setcitestyle{numbers} %hay que meter esto para que compile bien en arXive por un tema de la bibliografía
\icmltitlerunning{The Three-Ring Architecture: Governing Agents in the Era of On-Platform Organisations}

\makeatletter
\renewcommand{\printAffiliationsAndNotice}[1]{%
  \begingroup
  \renewcommand{\thefootnote}{\fnsymbol{footnote}}%
  \footnotetext[1]{%
    \textbf{Affiliations:} \\
    Sergio Álvarez-Teleña: Department of Computer Science, University College London, UK \\
    SciTheWorld, Spain\\
    Marta Díez-Fernández: SciTheWorld, Spain \\
    Correspondence to: sergio@scitheworld.com, marta@scitheworld.com
  }%
  \endgroup
}
\makeatother
\begin{document}

\twocolumn[
\icmltitle{The Three-Ring Architecture: Governing Agents in the Era of On-Platform Organisations}
\vspace{0.5em}
\begin{center}
Sergio Álvarez-Teleña\textsuperscript{1,2} \quad Marta Díez-Fernández\textsuperscript{2} \\
%Sergio Álvarez-Teleña \quad Marta Díez-Fernández \quad Natalia Cassinello \quad Ignacio Cervera\\
\end{center}
\vspace{0.5em}

\vskip 0.3in
]



\printAffiliationsAndNotice{}

\begin{abstract}
The current phase of enterprise AI deployment faces a structural failure: organisations are acquiring agentic capability without the infrastructure to govern it. The result is expected to reproduce the error of the first wave of AI deployment: decentralised intelligence without a federation layer leading to a $95\%$ project failure rate.\\

This paper formalises the Three-Ring Architecture as the governing infrastructure of the on-platform organisation. Ring 1 is the existing production architecture; Ring 2 is the M2 federation layer built on strategies-based agentic AI; Ring 3 is the LLM-based frontier intelligence layer. Ring 2 constitutes, in the technically exact sense, the operating system of the agentic enterprise - performing at the organisational level what a computing OS performs at the device level: resource abstraction, process coordination, permission enforcement, and a stable platform for compounding intelligence.\\

A central contribution is the formal distinction between Ring 2 and Ring 3 risk profiles. Strategies-based agents operate within a deterministic framework: their consequences are traceable, their permissions enforceable, their deviations recoverable. LLM-based agents introduce a categorically distinct risk: a non-deterministic actor whose deviations propagate through complex organisational systems without retrospective traceability. Ring 2 is not a useful addition - it is a necessary condition of control and compliance.\\

A further consequence: every improvement in LLM capability is a structural tailwind for this architecture. More capable non-deterministic actors produce larger consequences when they deviate. The governance requirement scales with capability. The architecture has been validated across a decade of deployment in financial services, government, procurement, and compliance.\\
\end{abstract}

\section{Introduction}
\label{section: Introduction}

\subsection{The Problem Being Named}
\label{subsection: The Problem Being Named}

The current discourse on enterprise AI is predominantly concerned with capability: the performance of frontier models, the autonomy of agentic systems, and the pace of improvement at the research frontier. These are legitimate objects of study. They do not, however, constitute the primary determinant of enterprise AI outcomes. The empirical record of deployment suggests that the governing constraint on enterprise AI transformation is not the capability of the models but the architectural conditions under which they are deployed.\\

The most consistently reported indicator of this constraint is the project failure rate. Across multiple independent assessments, approximately $95\%$ of enterprise AI initiatives fail to reach production (see MIT study\footnote{\href{https://mlq.ai/media/quarterly_decks/v0.1_State_of_AI_in_Business_2025_Report.pdf}{State of AI in Business 2025.}}). This figure exhibits a stability that is itself analytically significant. It has remained substantially unchanged across two structurally distinct deployment modalities: the first, characterised by human consultants deploying isolated algorithmic projects; the second, currently underway, characterised by LLM-based agents deployed at software speed. The persistence of the failure rate across modalities that differ in their technical substrate, deployment speed, and cost structure implies that its cause is not modality-specific. It is, we argue, architectural.\\

The common structural property of both deployment modalities is the absence of a federation layer: a governing architecture that coordinates distributed intelligence, enforces operational protocols, and provides the conditions under which locally generated capability can compound across the organisation rather than dissipate in isolation. In both waves, intelligence has been introduced into organisations in a decentralised and structurally ungoverned manner - project by project, department by department, vendor by vendor - without a shared protocol, a coordination mechanism, or an audit architecture. The consequences of this absence are well-documented empirically. The theoretical basis for why this absence produces failure at scale is less thoroughly developed in the existing literature, and constitutes one of the primary contributions of this paper.\\

\subsection{The Thesis}
\label{subsection: The Thesis}

This paper argues that the federation layer required for production-grade enterprise AI transformation constitutes, in the technically exact sense of the term, an operating system for the on-platform organisation. This claim is developed across three analytically distinct but interdependent dimensions.\\

The first dimension is functional. A computing operating system is defined not by its implementation but by the set of functions it performs: the abstraction of hardware complexity from the application layer, the coordination of concurrent processes, the enforcement of permissions and resource allocation, and the provision of a stable development platform on which applications can be built and compounded. We argue that the federation layer - designated Ring 2 in the Three-Ring Architecture introduced in this paper - performs an isomorphic set of functions at the enterprise level. The parallelism is functional rather than metaphorical, and its implications for how the federation layer should be designed, evaluated, and governed are non-trivial.\\

The second dimension is theoretical. The necessity of Ring 2 is not merely a practical observation derived from deployment experience. It is formally derivable from the properties of the system it governs. The enterprise, considered as an operational system, exhibits the characteristic properties of deterministic complexity: a large but bounded state space governed by defined protocols and interdependencies, in which local perturbations propagate non-linearly through the system. Within this framework, any actor operating at a node of the system - whether human or algorithmic - constitutes a potential sensitive initial condition whose local action may generate consequences of disproportionate magnitude at distant nodes. This is the class of problem that Ring 2 is designed to govern: not by eliminating complexity, but by bounding the state space within which any actor can operate and ensuring that deviations remain traceable and recoverable within a deterministic framework.\\

LLM-based agents, operating within Ring 3, introduce a risk profile that is categorically distinct from this class. They do not merely reproduce the sensitive-initial-condition problem within a deterministic system. They introduce non-determinism at the node level: their outputs bear a probabilistic rather than a functional relationship to their inputs, and their deviations - technically, hallucinations - propagate through the same complex organisational system without the retrospective traceability that deterministic complexity affords. The distinction, which we develop formally in Section \ref{section: The OS Analogy - Functional and Theoretical Foundations}, is between a butterfly operating within a deterministic system and a dice operating within the same system: the former generates consequences that are in principle traceable to their origin; the latter generates consequences whose origin is not recoverable from the system state. This distinction provides the theoretical grounding for a claim central to the paper: Ring 2 is not a governance preference within the Three-Ring Architecture. It is a necessary condition of its operational safety.\\

The third dimension is organisational. The destination of the Three-Ring Architecture is the on-platform organisation: a formal organisational state in which all working protocols, intelligence processes, and governance mechanisms operate within a single federated architecture, driven by business protocol rather than by technological capability. The on-platform organisation is not defined by the sophistication of its models or the autonomy of its agents. It is defined by the degree to which its intelligence compounds across departments and time within a governed architecture, rather than accumulating in isolated projects that do not interact. The transformation path toward this state - its diagnostic foundations, its governance instruments, and its economic structure - constitutes the operational content of the paper.\\

\subsection{Prior Work and Scope}
\label{subsection: Prior Work and Scope}
This paper extends the Algorithmization research programme developed by SciTheWorld over the preceding decade. The foundational technical constructs - including the Machine Theory of Agentic AI, the formal definitions of MAUs, MAEs, MAPs, and EPAs, and the distinction between M1 and M2 - are established in prior publications: principally \cite{Data MAPs} and \cite{BacktotheFuture}. These foundations are not reproduced here; readers unfamiliar with them are directed to those works.\\

The present paper extends that corpus in four directions. First, it formalises the Three-Ring Architecture as a complete and integrated framework, incorporating the theoretical grounding in deterministic complexity that prior publications did not develop explicitly. Second, it introduces the on-platform organisation as the formal destination of the architecture and characterises the transformation protocol - governed by the Three-Layer Company Model and the Ranked Transformation Agenda - as the organisational mechanism through which that destination is reached. Third, it reconstitutes the developer ecosystem argument with the business, rather than the technology function, as the primary driver of transformation. Fourth, it develops the market consequence of the OS framing with greater analytical precision than prior work, including the formal grounding for the proposition that improvements in LLM capability constitute a structural tailwind for Ring 2.\\

The contributions of this paper are theoretical, architectural, and economic. The underlying technology is described in prior publications. Validation is industrial: the architecture has been deployed across real organisations in operational environments over a decade, assessed by independent institutional actors, and recognised by relevant industry bodies. We consider this form of validation appropriate for the class of system under consideration, given that no controlled benchmark environment can reproduce the full operational complexity of the environments in which the architecture has been assessed.\\

\subsection{Structure of the Paper}
\label{subsection: Structure of the Paper}

Section \ref{section: The OS Analogy - Functional and Theoretical Foundations} establishes the OS analogy on functional and theoretical grounds, developing the deterministic complexity framework and the formal distinction between the butterfly and dice risk profiles. Section \ref{section: The Three-Ring Architecture} formalises the Three-Ring Architecture: its topology, five federation functions, and five architectural properties. Section \ref{section: Extreme-efficient Nations} introduces the Extreme-Efficient Nations diagnostic framework as the empirical foundation for transformation judgement. Section \ref{section: The Three-Layer Company Model and the Ranked Transformation Agenda} develops the Three-Layer Company Model and the Ranked Transformation Agenda as the governance instruments of the on-platform organisation. Section \ref{section: Algorithmization Training} presents Algorithmization training as the epistemic prerequisite of the transformation architecture. Section \ref{section: Two Compatible Technology Paths - Modular or Integral} develops the two compatible technology paths - modular and integral - and the governed allocation between them. Section \ref{section: Roles for Absorbing AI in the On-Platform Organisation} maps the role architecture for absorbing AI into the on-platform organisation, from the centre of excellence through the catalyser, the business, and the execution layer. Section \ref{section: Conclusions and Future Work} concludes, addresses the market consequence of the OS framing, and identifies directions for future work.\\

\section{The OS Analogy - Functional and Theoretical Foundations}
\label{section: The OS Analogy - Functional and Theoretical Foundations}

\subsection{ What an Operating System Actually Does}
\label{subsection:  What an Operating System Actually Does}

The operating system analogy is invoked frequently in technology discourse and applied with varying degrees of precision. For the purposes of this paper, it is necessary to establish a technically exact definition - one grounded in the functional properties of computing operating systems rather than in their cultural associations.\\

An operating system is not defined by its implementation, its commercial form, or its historical instantiation. It is defined by the set of functions it performs within a computational architecture. These functions are well-established in the computer science literature and can be stated with precision. First, resource abstraction: the OS mediates between the application layer and the underlying hardware, presenting applications with a uniform interface that conceals the complexity and heterogeneity of the physical substrate. Applications do not address memory registers, storage devices, or processing units directly; they address the abstractions the OS provides. Second, process coordination: the OS governs the concurrent execution of multiple processes, managing scheduling, preventing conflicts, and ensuring that simultaneously running applications do not produce inconsistent system states. Third, permission enforcement: the OS governs which processes can access which resources under which conditions, implementing a security and access control architecture that operates below the application layer and cannot be circumvented by any individual application. Fourth, platform provision: the OS provides a stable set of interfaces - system calls, APIs, and abstractions - on which applications can be built without requiring each application to solve the underlying hardware problem independently. This platform function is what enables a developer ecosystem: the stable substrate on which third parties can build applications that compound the value of the OS without modifying its core.\\

These four functions are not arbitrary. They are the precise set of functions required to make raw computational power productive at scale. A device with extraordinary processing capability but no operating system produces no value: it cannot allocate its resources, coordinate its processes, enforce its permissions, or support the construction of applications. The OS is not a convenience. It is the condition under which raw capability becomes productive infrastructure.\\

\subsection{The Enterprise AI Transition as an OS Moment}
\label{subsection: The Enterprise AI Transition as an OS Moment}
The structural parallel between the current enterprise AI transition and the personal computing transition of the early 1980s has been noted in prior work \cite{BacktotheFuture}. We develop it here with greater precision, and with explicit attention to the property that makes the parallel analytically useful rather than merely illustrative.\\

In the early 1980s, raw computational power became accessible to individuals and organisations at a price point that had previously been unthinkable. The processing capability of the available hardware was, relative to what had preceded it, extraordinary. What was absent was the infrastructural layer that would make that capability productive in practice. The introduction of DOS, and subsequently of Windows and Unix variants, provided that layer. These systems did not perform any of the tasks for which computing was ultimately used. They provided the conditions under which those tasks could be performed: abstracting hardware complexity, coordinating processes, enforcing permissions, and providing a stable development platform. The consequence was not a vertical product category but a horizontal infrastructure that restructured every sector of the economy over the following two decades.\\

The current enterprise AI transition exhibits the same structural pattern at the organisational level. Raw algorithmic intelligence - in the form of frontier language models, autonomous agents, and agentic pipelines - is becoming accessible to organisations at a cost that is declining rapidly and approaching, for many use cases, marginal zero. The capability of the available models is, relative to what preceded them, extraordinary. What is absent, across the overwhelming majority of organisations currently attempting to deploy this capability, is the infrastructural layer that would make it productive at enterprise scale: a layer that abstracts legacy complexity, coordinates intelligent processes, enforces permissions and compliance protocols, and provides a stable platform on which domain-driven intelligence can be built and compounded.\\

The empirical consequence of this absence - the $95\%$ project failure rate - is consistent with what would be predicted by the computing analogy. A device with extraordinary processing capability but no operating system produces nothing of value. An organisation with access to extraordinary model capability but no federation layer produces, at scale, the same outcome: ungoverned intelligence that cannot be coordinated, audited, or compounded.\\

\subsection{Deterministic Complexity and the Butterfly Problem}
\label{subsection: Deterministic Complexity and the Butterfly Problem}

The functional parallel between Ring 2 and a computing OS is, as argued above, precise. The theoretical grounding for Ring 2's necessity requires a further analytical step: an account of why the system it governs - the enterprise as an operational architecture - requires a governing layer of this kind in the first place. This account is provided by the framework of deterministic complexity.\\

The enterprise, considered as an operational system, exhibits the characteristic properties of a deterministic complex system. It consists of a large but finite set of actors - human and algorithmic - operating according to defined protocols and interdependencies, within a state space that is bounded by regulatory, contractual, and operational constraints. The system is deterministic in the formal sense: given a complete specification of its state at any moment and the actions taken by its actors, the subsequent state of the system is in principle derivable. It is complex in the formal sense: the interdependencies between actors and processes are sufficiently numerous and non-linear that the consequences of any local action cannot, in practice, be predicted from local information alone.\\

This combination of determinism and complexity generates a well-characterised class of governance problem. Small perturbations at one node of the system - a modification to a workflow protocol, a change to a data model, an alteration to a permission structure - may propagate through the network of interdependencies in ways that are non-linear, delayed, and disproportionate to the magnitude of the original perturbation. This is the property identified in the dynamical systems literature as sensitive dependence on initial conditions \cite{Deterministic_flow}, and its organisational manifestation has been documented in the systems thinking literature under various framings \cite{Mathematical_models}, \cite{Industrial_dynamics}.\\

For the purposes of this paper, we designate this the butterfly problem: any actor operating at a node of a deterministically complex system constitutes a potential sensitive initial condition. A developer independently modifying a protocol component, an analyst altering a data pipeline, a strategies-based agent adjusting a heuristic threshold - any of these may function as the butterfly whose local perturbation generates, through the system's non-linear interdependencies, consequences of disproportionate magnitude at distant nodes. The governing rationale for Ring 2 follows directly: an OS layer is required not because actors in the system are malicious or incompetent, but because the structural properties of the system ensure that local actions have non-local consequences. Through federation properties, the OS bounds the state space within which any actor can operate, enforces permission structures that contain propagation, and maintains the audit architecture that makes consequences traceable to their origins - rendering deviations recoverable even when they cannot be prevented.\\

This argument applies to any sufficiently complex deterministic system, and it provides the theoretical basis for the necessity of an OS layer in computing environments as well as in enterprise environments. It does not, however, fully characterise the governance problem posed by Ring 3. That problem is categorically distinct, and requires a separate analytical treatment.\\

\subsection{Non-Determinism at the Node Level: The Dice Problem}
\label{subsection: Non-Determinism at the Node Level: The Dice Problem}
LLM-based agents introduce a risk profile that is not reducible to the butterfly problem. The distinction is formal and carries significant implications for the governance architecture required.\\

A strategies-based agent operating within Ring 2 is, in the relevant sense, a deterministic actor. Given a defined input and a defined set of heuristics, it produces a defined output. Its behaviour is, in principle, fully specified by its design. The consequences it generates within the system may be non-linear and disproportionate - the butterfly problem applies - but they are traceable: given the system state and the agent's action, the causal chain connecting action to consequence is in principle reconstructable. This traceability is the property that makes Ring 2 governable: permission structures can be enforced, deviations can be detected, and audit trails can be maintained, because the agent's behaviour is a function of its inputs and its defined logic.\\

LLM-based agents do not share this property. Their outputs bear a probabilistic rather than a functional relationship to their inputs. The same input, presented to the same model at different moments or under different sampling conditions, may produce outputs that differ not merely in surface form but in substantive content and operational implication. More critically, LLM-based agents are subject to hallucination: the generation of outputs that are syntactically coherent and contextually plausible but factually or operationally incorrect, in ways that bear no recoverable relationship to any identifiable property of the input. Hallucination is not a failure mode that can be eliminated by improving the model; it is a structural property of the probabilistic generation process \footnote{Eloquently put, the $u$ in every $y = f(x) + u$ estimation, as explained in \cite{BacktotheFuture}}.\\

The governance problem this introduces is categorically distinct from the butterfly problem. We designate it the dice problem: a non-deterministic actor operating at a node of a deterministically complex system. The butterfly and the dice both generate perturbations that propagate non-linearly through the system. The butterfly's perturbation, however, is in principle traceable to a deterministic origin: the audit architecture of Ring 2 can, given sufficient instrumentation, reconstruct the causal chain from perturbation to consequence. The dice's perturbation has no deterministic origin to which it can be traced. The deviation - the hallucination - is a sample from a probability distribution, not the output of a defined function. Its consequences propagate through the same complex system, but the retrospective traceability that deterministic complexity affords is absent. The system state after a dice perturbation cannot be fully accounted for by reference to any recoverable initial condition.\\

This analysis has two direct implications for the Three-Ring Architecture. The first is that Ring 3 cannot govern itself: no application-level remedy - no prompt engineering, no output filtering, no self-consistency checking - resolves the fundamental governance problem posed by non-determinism at the node level within a deterministically complex system. The governing layer must operate at a level above the individual agent, with authority over the state space within which agents can act and the consequences they can propagate. This is Ring 2. The second implication concerns the relationship between LLM capability and governance requirements. As LLM-based agents become more capable, they are deployed in contexts of greater operational consequence, with access to more sensitive data and authority over more significant actions. A more capable dice, rolled within a complex deterministic system, generates perturbations of greater magnitude when it deviates. The governance requirement therefore scales with capability: every improvement in Ring 3 increases, rather than decreases, the necessity of Ring 2. We return to this point in Section \ref{section: Conclusions and Future Work}.\\

\subsection{Why Prior Enterprise Software Did Not Constitute an OS}
\label{subsection: Why Prior Enterprise Software Did Not Constitute an OS}

Given the functional definition of an operating system established in Subsection \ref{subsection:  What an Operating System Actually Does}, it is worth being precise about why prior enterprise software - despite its sophistication and its widespread adoption - did not constitute an OS in this sense. The argument matters because it establishes that the Three-Ring Architecture addresses a genuinely novel infrastructure problem, rather than extending or reframing existing solutions.\\

As noted a decade ago by the authors of \cite{Data MAPs} ERP systems performed specific tasks rather than governing the infrastructure for all tasks. They were, in the functional taxonomy of the OS analogy, applications: extraordinarily complex applications in their time, but applications nonetheless. Not only were they not federated but neither they abstracted anything below themselves - they required specialist integrators to connect them to the underlying production architecture, and that connection was specific to each deployment rather than general across deployments. Their permission and coordination functions were internal to the ERP's own processes rather than governing the broader enterprise intelligence layer. Their developer ecosystem was not a platform ecosystem in the OS sense; it was a certified integrator ecosystem whose scarcity reflected the depth of domain-specific knowledge required to extend the system.\\

SaaS platforms extended the ERP model to cloud delivery but preserved its fundamental character. Each platform performs a defined set of tasks within a defined domain. Each has an ecosystem of developers and integrators, but that ecosystem is structured around extending the platform's specific functionality rather than building arbitrary intelligence applications on a neutral substrate. The platform does not abstract the complexity of the enterprise from the application layer above it; it is the application layer.\\

Cloud infrastructure approaches the OS analogy most closely but operates at the wrong level of abstraction. It provides compute and storage infrastructure for software development broadly, and it does abstract hardware complexity in the sense defined in Subsection \ref{subsection:  What an Operating System Actually Does}. What it does not address is the specific problem of governing intelligence at the enterprise level: the coordination of intelligent agents, the federation of algorithmic strategies across departments, the integration of expert heuristics with statistical models, and the enforcement of compliance and IP protection across a distributed network of intelligent processes. Cloud infrastructure is the substrate on which Ring 2 can run. It is not Ring 2.\\

Ring 2 operates at the layer that none of these platforms reached: the governance of intelligence across the full operational fabric of the enterprise, in a manner that is federated, auditable, and compounding. It can gradually replace the layer below it. It governs the layer above it - and in doing so, it provides the condition under which the intelligence residing in Ring 3 can be made productive, safe, and compliant at enterprise scale.\\

\subsection{Section Conclusions}
\label{subsection: Section Conclusions}

The OS analogy, applied to Ring 2 of the Three-Ring Architecture, is functionally exact: Ring 2 performs, at the enterprise level, the same four functions - resource abstraction, process coordination, permission enforcement, and platform provision - that a computing OS performs at the device level. Prior enterprise software did not constitute an OS in this sense; it was application software that required specialist integrators and generated switching-cost moats rather than compounding intelligence.\\

The theoretical necessity of Ring 2 is established on two independent grounds. The first is the butterfly problem: the deterministic complexity of the enterprise as an operational system generates a class of governance requirement - bounding the state space within which actors operate, enforcing permissions, and maintaining audit traceability - that cannot be met at the application layer. The second is the dice problem: LLM-based agents introduce non-determinism at the node level within a deterministically complex system, generating a governance problem that is categorically distinct from the butterfly problem and for which no application-level remedy exists. Ring 2 is not a useful addition to an enterprise AI architecture that includes Ring 3. It is a necessary condition of that architecture's operational safety. The strength of this necessity scales with the capability of Ring 3: more capable non-deterministic actors, operating within a complex deterministic system, generate more consequential deviations and therefore require more robust governance.\\

\section{The Three-Ring Architecture}
\label{section: The Three-Ring Architecture}

\subsection{Overview and Motivation}
\label{subsection: Overview and Motivation}

The Three-Ring Architecture is the structural framework through which this paper formalises the governance of enterprise intelligence in the agentic era. It is composed of three analytically distinct but interdependent components, each defined by its functional properties, its boundary conditions, and its relationship to the governing layer at its centre.\\

The architecture is motivated by a diagnostic observation developed in Section \ref{section: Introduction}: the governing failure of enterprise AI deployment - across both the consultant-driven first wave and the agent-driven second wave - is not capability-related but structural. Organisations acquire intelligence capability and deploy it without a governing layer capable of coordinating, auditing, and compounding it across the operational fabric of the enterprise. The Three-Ring Architecture is the formal response to that structural failure. It does not prescribe which models to use, which workflows to automate, or which departments to transform first. It prescribes the architectural conditions under which any of those decisions can be made safely, governed coherently, and compounded over time.\\

The three rings are not a hierarchy of users. People interact at all three rings - with legacy applications in Ring 1, with strategies-based agents and new workflow interfaces in Ring 2, and with LLM-based tools in Ring 3. The rings are a hierarchy of infrastructure: each ring is defined by the properties of the systems it contains and the governance conditions under which those systems operate. AI may likewise enter at any ring - the ring of entry determines the governance conditions that apply, not the capability class of the system entering.\\

\begin{figure}[ht]
\vskip 0.1in
\begin{center}
\centerline{\includegraphics[width=\columnwidth]{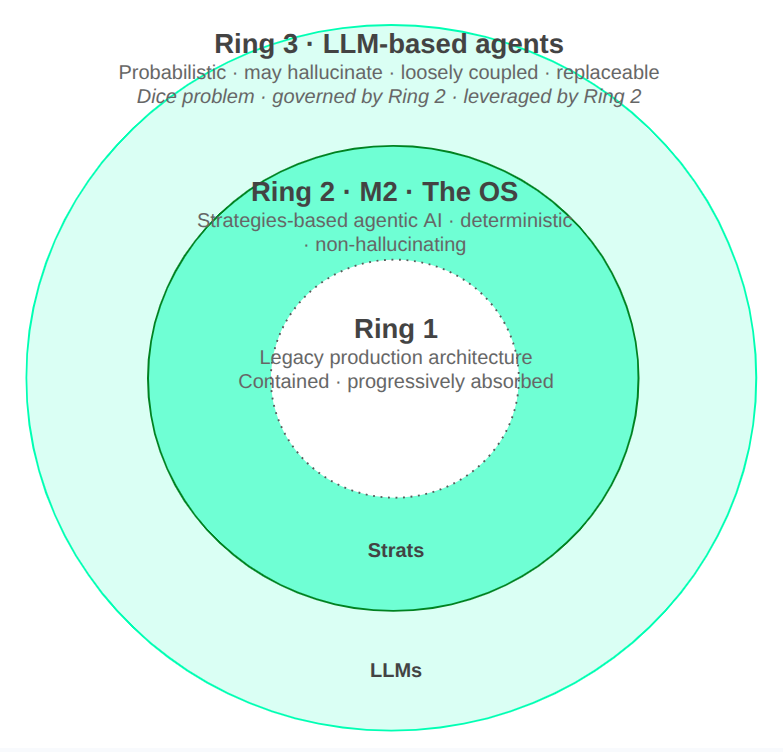}}
\caption{The Three-Ring Architecture. Ring 2 (solid boundary) is the OS: deterministic, governing, architecturally definitive.Ring 1 (inner dashed) is contained within Ring 2 and progressively absorbed. Ring 3 (outer dashed) is a loosely coupled satellite: probabilistic, replaceable, governed by but not contained within Ring 2. AI can enter at any ring.}
\label{fig: 3ring architecture}
\end{center}
\vskip -0.2in
\end{figure}

\subsection{Ring 1: The Legacy Production Architecture}
\label{subsection: Ring 1: The Legacy Production Architecture}

Ring 1 is the existing operational foundation of the enterprise: the ERPs, systems of record, regulated data stores, and production workflows that constitute the substrate on which the organisation currently runs. It is not a transitional artifact to be replaced. It is, in the precise architectural sense, the hardware layer of the enterprise OS - the physical and institutional infrastructure upon which everything above it depends.\\

Ring 1 has three properties that are analytically significant for the architecture. First, it is operationally indispensable: the organisation cannot cease to function while it transforms. The transformation must proceed while Ring 1 remains in production - changing the wheels while the car is moving, in the formulation we have maintained across prior work. Second, it is institutionally complex: Ring 1 systems encode years of regulatory compliance, contractual obligation, and operational knowledge that cannot be replicated quickly or safely from scratch. Third, it is progressively absorb-able: as Ring 2 matures, it builds new interfaces that mediate between users and Ring 1 systems, absorbing Ring 1 functions into the governed platform without disrupting the underlying systems. Over time, users cease to interact with Ring 1 directly for the functions Ring 2 has absorbed. Ring 2 agents interact with Ring 1 on their behalf.\\

The dashed boundary of Ring 1 in Figure \ref{fig: 3ring architecture} encodes this property precisely: Ring 1 is contained within Ring 2's architectural perimeter, but its boundary remains distinct. It is not dissolved into Ring 2 - it retains its own systems, data structures, and operational logic - but it is permeable to Ring 2's governance. The on-platform organisation is the limiting state in which Ring 1's boundary has been maximally permeated: Ring 2 governs all functions that Ring 1 previously governed directly, and Ring 1 persists only as the substrate Ring 2 runs on. There are further details on how to progressively let Ring 2 to take over Ring 1 in \cite{Data MAPs} where Extended Production Architectures (EPAs) unlock the change.\\

\subsection{Ring 2: The OS}
\label{subsection: Ring 2: The OS}
Ring 2 is the governing layer of the architecture - the strategies-based agentic AI platform designated M2, which performs at the enterprise level the same set of functions that a computing OS performs at the device level, as established in Section \ref{section: The OS Analogy - Functional and Theoretical Foundations}. It is the only ring with a solid boundary in Figure 1, reflecting its status as the architecturally definitive layer: deterministic, non-hallucinating, and non-negotiable as a condition of production-grade transformation.\\

Ring 2 performs three simultaneous functions that distinguish it from any prior enterprise software layer. The first is governance: Ring 2 governs the outputs generated by Ring 3 before they affect Ring 1, enforcing permissions, validating outputs against defined protocols, and maintaining the audit architecture that regulated enterprises require. The second is integration: Ring 2 federates Ring 1's legacy systems into the governed platform, providing the Extended Production Architectures through which new intelligence capabilities are connected to existing production infrastructure without disrupting it. The third is absorption: Ring 2 progressively builds new workflow interfaces that replace direct user interaction with Ring 1, absorbing Ring 1 functions into the on-platform architecture as the transformation advances.\\

The deterministic character of Ring 2 is not merely a technical property - it is the architectural prerequisite of its governance function. As established in Section \ref{section: The OS Analogy - Functional and Theoretical Foundations}, a non-deterministic layer cannot govern another non-deterministic layer. Ring 2's strategies-based agents operate on defined heuristics and produce outputs that are, in principle, fully traceable to their inputs. This is the butterfly problem, addressed: the consequences of any action within Ring 2 are bounded, traceable, and recoverable within a deterministic framework. It is precisely this deterministic character that makes Ring 2 capable of governing Ring 3, whose outputs are probabilistic and hallucinatory by design\footnote{ The error of any estimation from a probabilist model, that of the exploration from any Reinforcement Learning feature active and that from false friends, double meanings, etc implicit in written communication.}.\\

Last, Ring 2 is also the \textit{one-man band} of the architecture: a single governing layer that simultaneously integrates the foundation beneath it, governs the intelligence surface above it, builds new interfaces for users, mediates with legacy systems on their behalf, enforces compliance by architectural design, and compounds intelligence across departments and organisations - without external coordination and without centralising control in any single node.\\

\subsection{Ring 3: The Satellite}
\label{subsection: Ring 3: The Satellite}

Ring 3 is the LLM-based intelligence layer - the frontier language models, natural language interfaces, and agentic tools through which people and AI systems generate queries, code, and outputs. It is represented in Figure 1 as the outermost ring with a dashed boundary, encoding its fundamental architectural character: Ring 3 is a satellite. It orbits Ring 2. It is not contained within it.\\

The satellite characterisation is precise and carries three specific implications. First, Ring 3 is loosely coupled: it can be replaced, extended, or reconfigured without altering the governing architecture of Ring 2. A frontier model can be substituted for another, a new LLM tool can be introduced, written (or spoken) communication with agents can become outdated in terms of user experience, and Ring 2 continues to govern whatever Ring 3 produces. This replaceability is a structural property of the architecture, not a contingent feature of the current model landscape. Second, Ring 3 is probabilistic: its outputs bear a probabilistic rather than a functional relationship to its inputs, and its failure mode - hallucination - produces outputs that are syntactically coherent, contextually plausible, and operationally incorrect in ways that cannot be predicted in advance. This is the dice problem formalised in Section \ref{section: The OS Analogy - Functional and Theoretical Foundations}. Third, Ring 3 is externally governed: it does not govern itself. It requires Ring 2's deterministic layer to validate its outputs, enforce its permissions, and ensure that its probabilistic deviations do not propagate uncontrolled through Ring 1's operational infrastructure.\\

These three properties - loose coupling, probabilism, and external governance dependence - define Ring 3's architectural status. They also define the consequence of deploying Ring 3 without Ring 2: an ungoverned satellite whose probabilistic outputs interact directly with the deterministic complexity of the enterprise, generating the class of failure that the second wave of enterprise AI is currently producing at scale.\\

Ring 2's relationship to Ring 3 is not purely restrictive. Ring 2 leverages Ring 3 actively: deploying LLM-based agents within governed architectural contexts, using their generative capability for tasks where probabilistic outputs are appropriate and their consequences are bounded by Ring 2's governance. The two-way arrow in Figure 1 encodes this: Ring 2 governs downward from Ring 3, and Ring 2 leverages upward into Ring 3. The relationship is asymmetric - governance flows one way, leverage flows the other - but it is bidirectional.\\

\subsection{The Five Federation Functions}
\label{subsection: The Five Federation Functions}

Ring 2's governance of the Three-Ring Architecture is operationalised through five interdependent functions. Each function addresses a specific failure mode of decentralised AI deployment; together they constitute what we designate the federation layer - the complete set of capabilities required to make distributed intelligence productive, safe, and compounding at enterprise scale.\\

The first function is \textbf{aggregation}: the combination of outputs generated across multiple agents, departments, and systems into a coherent enterprise intelligence layer. Without aggregation, intelligence generated at one node of the organisation cannot inform the intelligence generated at another. Outputs accumulate in isolation, duplicating effort and preventing compounding. With aggregation, each new output becomes part of a shared intelligence substrate that grows with every contribution.\\

The second function is \textbf{coordination}: the orchestration of sequences, dependencies, and concurrent processes across the agent ecosystem. Without coordination, agents operating simultaneously may generate conflicting actions, produce inconsistent system states, or create race conditions that corrupt the operational integrity of Ring 1. With coordination, Ring 2 governs the sequencing of agent actions, managing concurrency and preventing conflicts across the full operational fabric.\\

The third function is \textbf{routing}: the direction of tasks to the appropriate agent, human, or system given the task's requirements, the agent's capabilities, and the compliance constraints that apply. Without routing, tasks reach the wrong agents, context is lost between handoffs, and accountability cannot be assigned. With routing, Ring 2 governs the flow of work through the architecture, ensuring that each task is processed by the most appropriate actor under the appropriate governance conditions.\\

The fourth function is \textbf{governance}: the enforcement of compliance, permissions, and audit requirements across the full agent ecosystem. Without governance, regulated enterprises cannot demonstrate the auditability of AI-assisted decisions, cannot enforce IP protection across distributed agent networks, and cannot satisfy the requirements of frameworks such as the EU AI Act and DORA. With governance, Ring 2 produces compliance by architectural design rather than post-hoc audit - the platform is the compliance factory.\\

The fifth and master function is \textbf{federation}: the enablement of cross-company and cross-industry intelligence sharing with IP protection preserved at every node. Without federation, intelligence compounds only within the boundaries of the individual organisation. With federation, a risk management protocol developed by one organisation can inform the architecture of another - without either organisation exposing its proprietary data or logic to the other. Federation is the function that transforms the Three-Ring Architecture from an organisational tool into an ecosystem: a governed network across which intelligence accumulates at a rate that no individual organisation could achieve alone.\\

\subsection{The Five Architectural Properties}
\label{subsection: The Five Architectural Properties}

Beyond its functional layer, Ring 2 carries five architectural properties that regulated enterprises require and that Ring 3 alone cannot provide.\\

The first is \textbf{ease of integration}: Ring 2's Extended Production Architectures allow co-existence with Ring 1 legacy systems from day one of deployment. The structural transformation does not require Ring 1 to be replaced or suspended. It proceeds in parallel with Ring 1's continued operation, absorbing Ring 1 functions progressively as the platform matures. This property is what makes the changing-wheels-while-driving formulation operationally achievable rather than merely aspirational.\\

The second is \textbf{IP protection by design}: Ring 2 operates on the client's own infrastructure. Proprietary models, data, and actions remain on the client's servers. The SciTheWorld layer is removable at any time. No proprietary information transits through external vendor infrastructure as a condition of the platform's operation. This property is not a security feature added to Ring 2 - it is a consequence of its federated, on-premises architecture.\\

The third is \textbf{control}: Ring 2 governs agent behaviour through architectural logic rather than prompt engineering. Agent actions are orchestrated by defined strategies and heuristics, not improvised from probabilistic generation. This property is what distinguishes Ring 2's deterministic agents from Ring 3's probabilistic ones - and it is the property that makes Ring 2's governance of Ring 3 technically coherent rather than merely aspirational.\\

The fourth is \textbf{auditability}: Ring 2's architecture produces decision trails by design. Every agent action is traceable to its architectural source - the strategy, the heuristic, the permission condition that authorised it. This property directly satisfies the auditability requirements of the EU AI Act and equivalent regulatory frameworks, which regulate systems rather than models and require the traceability of decisions rather than the explainability of weights.\\

The fifth is \textbf{traceability}: Ring 2's micro-architecture units - MAUs - make every agent action traceable to its architectural origin. When something fails in production, the failure is not a black box. It is recoverable to a defined component of the architecture, diagnosable, and correctable without system-wide disruption. This property is the operational expression of the deterministic complexity argument made in Section \ref{section: The OS Analogy - Functional and Theoretical Foundations}: in a deterministic system, consequences are traceable to their origins. Ring 2 preserves that traceability as an architectural guarantee.\\

\subsection{Section Conclusions}
\label{subsection: Section 3 Conclusions}

The Three-Ring Architecture formalises the structural response to the governance failure of enterprise AI deployment. Ring 1 is the legacy foundation - necessary, progressively absorbed, contained within Ring 2's perimeter but retaining its own operational identity. Ring 2 is the OS - deterministic, non-hallucinating, governing both the integration of Ring 1 and the outputs of Ring 3 through five interdependent federation functions and five architectural properties that regulated enterprises require. Ring 3 is the satellite - probabilistic, loosely coupled, replaceable, and dependent on Ring 2's governance for its safe operationalisation within Ring 1's deterministic infrastructure.\\

The architectural consequence of this structure is precise: the capability of Ring 3 is not the limiting factor on enterprise AI transformation. The governance provided by Ring 2 is. Every improvement in Ring 3 capability increases the necessity of Ring 2's governance - more capable probabilistic actors, operating within complex deterministic systems, require more robust governance, not less. The Three-Ring Architecture is not a response to the current state of LLM capability. It is a response to the structural properties of the system those models are being deployed into - properties that do not change as the models improve.\\

A further consequence of this framing concerns the current market configuration of enterprise AI deployment, which exhibits a structural pattern that the Three-Ring Architecture renders analytically legible. The dominant commercial actors in the current wave of enterprise AI are Ring 3 vendors: frontier model providers and the agentic tools built on top of them. Their media presence, investment capture, and organisational mindshare are disproportionate to their architectural completeness. In a significant and growing number of enterprise deployments, Ring 3 vendors are connecting directly to Ring 1 systems - bypassing Ring 2 entirely. This pattern is not coincidental. Transformation is becoming a byproduct of LLM capability rather than the object of independent architectural design: a pivot toward enterprise relevance driven by commercial positioning rather than by the systematic study of what production-grade transformation requires. This is not optimal design arrived at through first-principles research. It is product-market fit sought after the product already exists - and the architecture that results bears the structural imprint of that inversion.\\

This dynamic was identified and formally characterised in prior work. In \cite{Data MAPs}, we documented that the ecosystem of technology providers has largely delivered siloed projects with little to no architecture synergies exploitation - a pattern that produces organisations that are effective at reaching isolated digital impacts but structurally inefficient, facing higher costs and architectures that slow their evolution significantly. The mechanism behind this pattern is the monopolistic incentive structure of the Digital Dis-Economy: a provider that controls a siloed node of the enterprise architecture preserves its position as a necessary intermediary at that node, whereas a neutral governing architecture would reduce that dependency by federating intelligence across nodes without privileging any single provider. Ring 3 vendors operating without Ring 2 reproduce this pattern at greater speed and scale: each vendor occupies a node, delivers a project, and generates the spaghetti-like structures that the paper identified as the characteristic output of non-platformed transformation.\\

The Three-Ring Architecture was not designed in response to this market configuration. It predates it. Its origins lie in algorithmic trading - the most demanding federated agent environment in existence, and one in which the consequences of ungoverned probabilistic actors interacting directly with deterministic production systems were understood and addressed two decades before the current wave of LLM deployment. The architecture that emerged from that environment is aseptic in the precise sense: its design was not shaped by the commercial pressures, positioning constraints, or capability narratives of Ring 3 vendors. It was optimised, from first principles and over a decade of production deployment, for the single architectural problem that Ring 3 vendors cannot solve for themselves - and that enterprises cannot solve without it. The completeness of the algorithmic architecture, as the paper established, requires simultaneous depth in business expertise, machine learning, and technology architecture: a combination that no Ring 3 vendor, optimised for model capability, is currently structurally positioned to provide.\\

\section{The Diagnostic Foundation - Extreme-Efficient Nations and True Transformation Judgement}
\label{section: Extreme-efficient Nations}

\subsection{The Problem with Limited Self-Assessment}
\label{subsection: The Problem with Limited Self-Assessment}

The decision to undergo enterprise AI transformation is, in most organisations, preceded by some form of internal diagnostic: an inventory of existing systems, an assessment of current AI maturity, a survey of departmental needs, or a consultation with external advisors. These diagnostics are not without value. They surface manifest problems, establish a baseline, and provide the organisational vocabulary through which transformation can be discussed. Their limitation is structural rather than methodological: they are bounded by the organisation's own frame of reference - and, critically, by the frame of reference of the advisors the organisation has already engaged.\\

This second constraint is frequently underestimated. External consultants and technology providers approached for diagnostic input are not neutral instruments of organisational self-knowledge. They arrive with pre-existing solutions, pre-existing client relationships, and a structural incentive to frame organisational problems in ways that their available offerings address. The organisation that consults advisor A, who proposes solutions to pains a, b, c, and d, and advisor B, who proposes solutions to pains a, c, e, and f, will construct its transformation agenda from the union of those letter sets - without any mechanism for determining whether pains g, h, i, and the rest up to z (or even alpha, omega…) are more impactful, more urgent, or more strategically significant than any of the pains its advisors have named. The agenda is selected from a menu the organisation did not design, without knowledge of how many items are absent from it. This myopia in priority selection is a structural property of advisor-mediated diagnostics, not an aberration.\\

An organisation assessing its own AI maturity under these conditions cannot know what it does not know. Its blind spots are, by definition, invisible to it. The pains it articulates are the ones it has already named - the problems that have risen to the level of conscious organisational attention through internal deliberation or external prompting. The pains it does not articulate are frequently more consequential: the latent dysfunctions that have not yet been named as problems, the structural inefficiencies that have been normalised through years of accommodation, the competitive vulnerabilities that are only visible when measured against the performance of comparable organisations operating under different conditions.\\

This epistemic limitation has a direct consequence for transformation planning. An organisation that prioritises its transformation agenda based exclusively on internally generated diagnostics - whether self-administered or standard-advisor-mediated - will systematically underweight the problems it cannot see and overweight the problems it has already begun to address or that its existing advisors are positioned to solve. The result is a transformation agenda that optimises for the organisation's current self-model and its existing advisory relationships rather than for its actual competitive position. The gap between the two is precisely the gap that a cross-sector diagnostic framework is designed to close.\\

\subsection{The Extreme-Efficient Nations Framework}
\label{subsection: The Extreme-Efficient Nations Framework}

The Extreme-Efficient Nations framework was developed by SciTheWorld as the diagnostic foundation of a macro project - first of its kind - conducted in collaboration with the Ministry of Economy of Spain, the Instituto de Empresa Familiar, the Real Instituto Elcano, and the Instituto de Crédito Oficial. The project's first step was precisely this: the design and administration of a diagnostic instrument capable of assessing organisational and national AI transformation readiness at scale, across a sufficiently large and diverse population of organisations to generate cross-sector analytical conclusions that no single-organisation or single-sector assessment could produce.\\

The governing principle of the EEN framework is the homogenisation of risk-reward assessments across organisations and sectors. By subjecting a large and diverse population of organisations to a common diagnostic instrument, the framework generates two analytically distinct but interdependent outputs. The first is a deep internal cross-department prioritisation: a structured ranking of the organisation's own pains, opportunities, and transformation prerequisites, produced through a diagnostic process that operates across all functions simultaneously rather than department by department. The second is a broad cross-sector benchmark: a calibration instrument that allows the organisation to locate its own position - across every dimension the diagnostic measures - relative to the full population of organisations that have passed through the same instrument. The precision of this dual output exceeds anything a partial, advisor-mediated, or internally generated assessment can provide, for the structural reasons established in Subsection \ref{subsection: The Problem with Limited Self-Assessment}.\\

The framework's diagnostic instrument is a structured questionnaire administered across organisational functions - covering technology infrastructure, data governance, AI deployment maturity, workforce readiness, regulatory compliance posture, and transformation culture. The questionnaire is calibrated to surface both manifest and latent pains: it includes direct questions about identified problems and indirect probes designed to reveal dysfunctions that respondents have not yet articulated as problems. Black-ops methodological elements - indirect indicators of real predisposition toward transformation, beyond what conventional surveys capture - are incorporated to address the well-documented gap between stated and revealed organisational attitudes toward change.\\

The framework operates under strict regulatory confidentiality protocols. The authors are, among other things, AI ISO committee members - an association that formalises best praxis on the field. Individual organisational responses contribute to the aggregate benchmark without any individual response being attributable to its source. This confidentiality architecture is not merely a procedural requirement. It is the condition under which the cross-sector benchmark acquires analytical credibility: organisations contribute honestly to an instrument whose aggregate outputs cannot be traced back to them individually, producing a benchmark that reflects actual organisational states rather than the curated narratives that non-confidential assessments tend to generate.\\

\subsection{The Scanning Process}
\label{subsection: The Scanning Process}

The EEN diagnostic is not administered as a conventional audit. It is orchestrated as a structured elicitation process, coordinated by a member of the SciTheWorld team who works across departments to ensure that the diagnostic captures the full operational complexity of the organisation rather than the curated narrative that any single function would present in isolation.\\

The scanning process operates on two simultaneous levels. The first is the level of governance: identifying which existing AI and data developments, processes, and tools are compatible - or can be made compatible - with the federated architecture to be deployed, and which require revision before integration. This level of the diagnostic produces a structured inventory of the organisation's current algorithmic assets and their governance status - a direct input to the prioritisation protocol developed in Section \ref{section: The Three-Layer Company Model and the Ranked Transformation Agenda}.\\

The second level is that of digital maturity assessment. The most common risk the diagnostic must identify is not the absence of tools but their structural underutilisation: new technological capabilities fitted within pre-existing workflows rather than used as a lever to redesign those workflows from first principles. This pattern is systematic when teams are unaware of the real limits of the available technology - and it is, precisely, one of the most revealing indicators of transformational maturity. An organisation that has deployed AI extensively but has not redesigned its workflows around that deployment is not more mature than one that has deployed less but redesigned more. The diagnostic is calibrated to surface this distinction.\\

Two further analytical dimensions are embedded in the scanning process that warrant explicit treatment.\\

The first is the identification of unknown but latent pains - those that do not emerge in standard diagnostic conversations because they have not yet been articulated as problems at the organisational level. A paradigmatic example is the systematic turnover of Generation Z technical talent, whose impact on the continuity of operational knowledge and process governance is frequently underestimated until it becomes a structural risk. The scanning process is designed to surface these latent pains through indirect probes and behavioural indicators that conventional interviews and surveys do not capture.\\

The second is the determination of the organisation's data model ontology: whether existing data structures follow an object-oriented logic - defining key entities such as employees, clients, and suppliers by their attributes and possible actions within the organisation - which constitutes the ontological prerequisite for the kind of algorithmic architecture Ring 2 requires. Organisations whose data models do not yet follow this logic require a prior architectural intervention before the full Three-Ring Architecture can be deployed at scale. The diagnostic identifies this requirement early, preventing it from emerging as a production bottleneck at a later and more costly stage of the transformation.\\

A further element of the scanning process concerns the distinction between what constitutes a must - a necessary condition for operations or regulatory compliance - and what is a nice to have. Within the latter category, the diagnostic further distinguishes between improvements that represent an expected advance for the entire industry - and therefore generate no differential competitive advantage - and developments that constitute proprietary progress susceptible to protection and capitalisation. This is the right-to-play versus right-to-win distinction that structures the Three-Layer Company Model developed in Section \ref{section: The Three-Layer Company Model and the Ranked Transformation Agenda}. The scanning process is the empirical instrument through which this distinction can be drawn with analytical rigour rather than strategic intuition.\\

\subsection{The Sector Benchmark}
\label{subsection: The Sector Benchmark}

The primary output of the EEN diagnostic, beyond the individual organisational assessment, is the sector benchmark: a cross-sector ranking of organisational pain profiles, AI maturity levels, and transformation readiness indicators, produced with anonymity guarantees and enriched, where sector population is sparse, by adjacent-sector comparators with comparable size and operational profiles.\\

The sector benchmark provides something that no internal diagnostic and no advisor-mediated assessment can supply: an empirically grounded, cross-sector view of where the organisation actually stands relative to comparable organisations across every dimension the diagnostic measures. This is not a competitive ranking in the market sense. It is a calibration instrument: it tells the organisation which of its pains are common across its sector - and therefore generate no differential advantage if addressed - and which are idiosyncratic to its situation, representing either proprietary vulnerabilities that require urgent attention or proprietary opportunities that competitors have not yet identified.\\

The sector benchmark also performs a second analytical function that is strategically significant: it reveals the transformation trajectory of the sector as a whole. An organisation that understands not only where it stands today but where its sector is moving - which capabilities are becoming right-to-play requirements, which are still right-to-win opportunities, and at what rate the frontier is advancing - is in a structurally superior position to sequence its transformation investments than one operating without this reference frame.\\

This distinction - between right-to-play improvements and right-to-win developments - is one of the most actionable outputs of the diagnostic process. It is the distinction that separates a transformation agenda that makes the organisation competitive from one that merely makes it compliant. The sector benchmark is the empirical foundation on which this distinction can be drawn with analytical precision, and it is a foundation that no organisation can construct for itself from internally generated data alone.\\

\subsection{From Diagnosis to Agenda}
\label{subsection: From Diagnosis to Agenda}

The output of the EEN diagnostic process is not a report in the conventional sense. Reports are consumed, filed, and superseded. The EEN output is the empirical foundation of the transformation agenda - the structured input to the prioritisation protocol that determines the sequence, depth, and resource allocation of every transformation project the organisation will undertake. Its function is constitutive rather than documentary: it does not describe the organisation's current state for archival purposes; it provides the empirical substrate from which the transformation agenda is built.\\

This distinction has a direct consequence for the analytical integrity of the agenda. A transformation agenda derived from internal consensus is vulnerable to the political economy of the organisation: it will tend to reflect the priorities of the most articulate or most powerful departments, the solutions of the most established advisory relationships, and the problems that are most visible rather than most consequential. A transformation agenda derived from a cross-sector empirical benchmark is structurally more resistant to these distortions: it provides an externally grounded ranking of pains and opportunities whose analytical basis is independent of the organisation's internal political economy and existing advisor relationships.\\

The EEN diagnostic performs a function in the transformation process that is analogous to the function of an independent valuation in capital allocation: it provides the external reference point against which internally generated assessments can be calibrated, corrected, and prioritised. The precision of this calibration is a direct function of the population size and diversity of the organisations that have passed through the same diagnostic instrument - a property that compounds over time as the framework accumulates observational depth.\\

A further consequence of the framework's aggregation logic operates at the national rather than the organisational level. The cross-sector benchmark, accumulated across a sufficiently large and diverse population of organisations, constitutes an empirical signal of national economic transformation readiness that is methodologically distinct from - and analytically superior to - the indicators typically available to economic policy authorities. Macroeconomic policy directed at enterprise AI transformation has historically been designed top-down: derived from theoretical models, expert opinion, or aggregate statistics that do not capture the granular operational reality of the organisations the policy is intended to affect. The EEN framework inverts this logic. By aggregating a bottom-up signal constituted from real organisational diagnostic states across sectors, it provides the ministry of economy - or equivalent national authority - with an empirically grounded basis for designing economic policies that address measurable, operationally defined problems rather than theoretically derived ones. The policy instrument is calibrated to the actual distribution of organisational needs rather than to a pontification of what those needs are assumed to be. This is the methodological foundation of the Extreme-Efficient Nations thesis at the national level - and it is a direct consequence of the same diagnostic architecture that governs transformation prioritisation at the organisational level.\\

\subsection{Section Conclusions}
\label{subsection: Section 4 Conclusions}

The EEN diagnostic framework addresses the structural limitations of self-assessment and advisor-mediated diagnostics by providing a cross-sector empirical benchmark against which any organisation can locate its actual transformation position with a precision that partial assessments cannot achieve. Its scanning process surfaces both manifest and latent pains, assesses digital maturity at the workflow level rather than the tool level, identifies the ontological prerequisites of federated architectural deployment before they become production bottlenecks, and draws the right-to-play versus right-to-win distinction that structures the transformation agenda. Its sector benchmark calibrates the organisation's pain profile against a population-level distribution - revealing both the urgency of specific interventions and the trajectory of the sector as a whole.\\

The diagnostic framework is, in this sense, the epistemological prerequisite of the transformation protocol developed in Section \ref{section: The Three-Layer Company Model and the Ranked Transformation Agenda}. A prioritisation algorithm applied to an incomplete or distorted problem set will produce an optimised agenda for the wrong problems. The EEN framework is the instrument through which the problem set is constituted with sufficient completeness and cross-sector calibration to make the prioritisation algorithm analytically productive. At the national level, the same aggregation logic that calibrates individual transformation agendas constitutes the empirical substrate of evidence-based economic policy - inverting the conventional top-down policy design logic and grounding national economic strategy in the operationally defined needs of the enterprise population it is designed to serve. Section \ref{section: The Three-Layer Company Model and the Ranked Transformation Agenda} develops the prioritisation protocol through which diagnostic outputs are converted into a governed, compounding transformation agenda.\\

\section{The Three-Layer Company Model and the Ranked Transformation Agenda}
\label{section: The Three-Layer Company Model and the Ranked Transformation Agenda}

\subsection{The Governance Problem It Solves}
\label{subsection: The Governance Problem It Solves}

The diagnostic framework developed in Section \ref{section: Extreme-efficient Nations} produces a comprehensive, cross-sector-calibrated map of the organisation's transformation position: its pain profile, its digital maturity level, its ontological prerequisites, and the sector benchmark against which its priorities can be located. The analytical value of this map is not self-executing. A comprehensive problem inventory, without a governed prioritisation instrument, does not produce a transformation agenda. It produces a list - and lists, in organisations of sufficient complexity, are consumed by the political economy of the organisation rather than acted upon in order of genuine strategic value.\\

The governance problem that Section \ref{section: The Three-Layer Company Model and the Ranked Transformation Agenda} addresses is therefore distinct from the epistemic problem addressed in Section \ref{section: Extreme-efficient Nations}. The epistemic problem was: how does the organisation know what its real problems are, independent of the distortions introduced by internal self-assessment and advisor-mediated diagnostics? The governance problem is: given a comprehensive and calibrated problem inventory, how does the organisation convert it into a sequenced, resourced, and compounding transformation agenda that is resistant to the political economy of the organisation, transparent to its board, and self-correcting over time?.\\

The answer developed in this section is the Three-Layer Company Model and its operational instrument, the Ranked Transformation Agenda. Together they constitute the governance architecture of the on-platform organisation: the layer that sits between the diagnostic foundation of Section \ref{section: Extreme-efficient Nations} and the architectural deployment of the Three-Ring Architecture, converting empirical inputs into governed action.\\

\subsection{The Three-Layer Company Model}
\label{subsection: The Three-Layer Company Model}

The Three-Layer Company Model is a strategic classification framework that organises every transformation project - and every organisational capability - into one of three analytically distinct categories, defined by their relationship to competitive advantage.\\

The first layer is \textbf{Tech Must}: the set of capabilities that constitute necessary conditions for the organisation's continued operation or regulatory compliance. These are not strategic choices. They are operational prerequisites whose absence would expose the organisation to regulatory sanction, operational failure, or competitive exclusion. In the context of the Three-Ring Architecture, Tech Must projects typically concern Ring 1 integrity - the maintenance, governance, and security of the legacy production architecture - and the regulatory compliance properties of Ring 2, including EU AI Act and DORA requirements. Tech Must projects are not prioritised against each other on the basis of strategic value; they are prioritised on the basis of urgency and risk of non-compliance.\\

The second layer is \textbf{Right to Play}: the set of capabilities that the industry expects of any competitive participant but that generate no differential advantage if present. Right to Play capabilities are the minimum viable position in the competitive landscape: their absence constitutes a liability, but their presence does not constitute an advantage. In the context of enterprise AI transformation, Right to Play capabilities are those that are becoming standard across the sector - visible in the sector benchmark as the median capability level of comparable organisations. Investing in Right to Play capabilities is necessary but not sufficient for competitive differentiation. The strategic risk is over-investing in them at the expense of Right to Win development.\\

The third layer is \textbf{Right to Win}: the set of capabilities that are proprietary, protectable, and competitively differentiating - developments that comparable organisations have not yet achieved, that are not yet visible in the sector benchmark as standard capabilities, and that constitute genuine competitive advantage susceptible to capitalisation. Right to Win projects are the highest-value projects in the transformation agenda, but they are also typically the most complex, the most resource-intensive, and the most dependent on the architectural foundations that Tech Must and Right to Play projects establish. The Three-Layer Company Model governs the sequencing of investment across all three layers - ensuring that Right to Win ambition is not pursued at the expense of the Tech Must and Right to Play foundations it depends upon, and that Right to Play investment does not consume the resources required for Right to Win development.\\

The Three-Layer Company Model performs two high-value functions simultaneously. The first is strategic: it gives the board a principled framework for evaluating the transformation portfolio that is independent of departmental lobbying and advisory solution sets. The second is communicative: it provides a shared vocabulary through which business leaders, technology leaders, AI leaders, and transformation leaders can discuss investment priorities without the conceptual friction that arises when these functions approach prioritisation from incommensurable frames of reference.\\

\subsection{The Ranked Transformation Agenda}
\label{subsection: The Ranked Transformation Agenda}

The Ranked Transformation Agenda is the operational instrument through which the Three-Layer Company Model is applied to the organisation's specific transformation context. It is not a static prioritisation exercise - a one-time ranking of projects produced at the outset of transformation and consulted thereafter. It is a dynamic, continuously updated prioritisation algorithm that governs the allocation of transformation resources in real time as the organisation's context evolves, new projects emerge, and completed projects compound the value of those that follow.\\

The RTA is governed by four analytical inputs. The first is the EEN diagnostic output: the cross-sector-calibrated pain inventory and sector benchmark established in Section \ref{section: Extreme-efficient Nations}, which provides the empirical foundation for the initial ranking. The second is quantitative impact assessment: the estimated value of addressing each identified pain, measured in terms of the specific operational and financial consequences for the organisation rather than generic industry benchmarks. The third is calibrated heuristics: the expert judgement of domain professionals and transformation leaders, incorporated into the ranking algorithm as weighted inputs that reflect operational knowledge the quantitative assessment cannot capture. The fourth is game theory: the anticipation and modelling of possible deviations by the actors involved in project execution - departments that may resist prioritisation decisions that reduce their resource allocation, vendors that may advocate for project sequences that favour their solution portfolios, and technical teams that may systematically underestimate execution complexity. The RTA incorporates these behavioural dynamics as explicit inputs to the ranking algorithm rather than treating them as noise to be managed outside the prioritisation process.\\

The RTA also incorporates a fast consultancy layer: agile external feedback on estimated costs, execution timelines, and implementation challenges that enriches the risk-benefit evaluation of each project with operational insight that the organisation's internal teams may not possess. This feedback is incorporated as an input to the algorithm rather than as a override of its outputs: the consultancy layer informs the ranking but does not replace it.\\

The governing KPI of the RTA is Time-to-Production: the elapsed time between the initiation of a transformation project and the delivery of value in production. TTP is not a conventional project management metric. It is, as established across prior work in this programme, the most reliable empirical predictor of transformational success: every week without value in production is organisational capital that erodes, and the speed of reaching production is the primary determinant of whether the transformation compounds or stalls. The RTA is calibrated to minimise TTP across the full project portfolio - not for any individual project in isolation, but for the portfolio as a whole, taking into account the dependency structures between projects and the synergies that shared architectural foundations generate.\\

\subsection{Quick Wins and Structural Projects}
\label{subsection: Quick Wins and Structural Projects}

The RTA governs the sequencing of two analytically distinct project types that must be managed in explicit relationship to each other throughout the transformation.\\

Quick wins are projects whose impact is immediate, whose execution complexity is low, and whose value can be delivered in production within a short time horizon. Their strategic function in the transformation is not primarily their direct impact - though that impact is real and valuable. It is their organisational function: quick wins establish the credibility of the transformation methodology, demonstrate to the board that investment is producing visible value, and generate the organisational momentum that sustains commitment to the more complex structural projects that follow. The RTA governs the selection of quick wins with this organisational function in mind: the optimal quick win is not the easiest project but the project whose rapid delivery most effectively sustains the conditions for structural transformation.\\

Structural projects are the most complex and highest-value components of the transformation: the projects that establish the architectural foundations on which the full Three-Ring Architecture operates, and on which subsequent projects compound. They are characterised by longer time horizons, higher resource requirements, greater dependency on prior architectural foundations, and non-trivial prioritisation logic that the RTA must govern actively to prevent scope expansion, resource dissipation, and the indefinite deferral that complexity tends to produce in the absence of a governed prioritisation instrument.\\

The critical analytical point about structural projects - one that distinguishes the Three-Ring Architecture approach from the conventional enterprise transformation model - is that they do not start from scratch. Ring 2's architecture is already built and validated. The structural component of deployment consists of tuning that architecture to the organisation's specific context, not constructing it. This distinction has a direct consequence for the TTP of structural projects: they reach production faster than equivalent projects built from first principles, and their cost is distributed across the organisation's full project portfolio through the synergies that a shared architectural foundation generates.\\

\subsection{Synergies and the Compounding Effect}
\label{subsection: Synergies and the Compounding Effect}

The most analytically significant property of the RTA - and the property that most clearly distinguishes on-platform transformation from the isolated project model that preceded it - is its explicit governance of synergies between projects.\\

In the isolated project model documented in \cite{Data MAPs}, innovation within enterprises occurs in isolated pockets: technology providers deliver siloed projects with little to no architecture synergies exploitation, producing organisations that are effective at reaching isolated digital impacts but structurally inefficient. Each project bears its full development cost independently; no project benefits from the architectural foundations established by prior projects; and the organisation's transformation portfolio accumulates as a collection of disconnected investments rather than as a compounding architecture.\\

The RTA inverts this model. Because all projects are deployed on the shared architectural foundation of Ring 2, every project contributes components - heuristics, data models, workflow protocols, compliance structures - that are natively inheritable by subsequent projects. The cost of each subsequent project is therefore lower than the cost of the preceding one, and its ROI is higher, because it builds on an architectural foundation that prior projects have already established and validated. This is the compounding effect of the on-platform model: transformation accelerates as the platform matures, rather than accumulating the friction and complexity that isolated project portfolios generate over time.\\

The RTA makes this compounding effect explicit and governable. It identifies, for each candidate project, the synergies it would generate for subsequent projects and the synergies it would inherit from prior ones - incorporating these into the ranking algorithm as explicit inputs. A project with modest direct impact but high synergy generation may rank above a project with higher direct impact but low synergy generation, because the former accelerates the entire subsequent portfolio while the latter does not. This synergy-weighted prioritisation is the mechanism through which the RTA produces a transformation agenda that compounds rather than merely accumulates.\\

\subsection{The Fractal Governance Principle}
\label{subsection: The Fractal Governance Principle}

A property of the Three-Layer Company Model and the RTA that warrants explicit treatment is what we designate the fractal governance principle: the governance of the transformation is itself executed on-platform from the first day of deployment.\\

The prioritisation of resources, the management of project reporting, the measurement of deviations against KPIs, and the production of board-level overviews of the transformation portfolio - all of these governance functions are executed within the Ring 2 architecture rather than through external tools, spreadsheets, or advisory presentations. This is not merely an efficiency property. It is an architectural consistency property: the platform that governs the organisation's intelligence also governs the transformation that is building it. The organisation learns to use the platform by using it to govern its own transformation - generating a feedback loop between transformation governance and platform maturity that accelerates both simultaneously.\\

The fractal nature of this principle manifests at multiple levels. At the project level, each project's reporting, deviation measurement, and KPI tracking are executed on-platform. At the portfolio level, the RTA ranking algorithm and its inputs are executed on-platform. At the board level, the Three-Layer Company Model overview - the strategic view of the transformation portfolio across Tech Must, Right to Play, and Right to Win layers - is produced self-drawn by the platform, without dependence on external intermediation. The organisation develops the capacity to read its own transformation autonomously, rather than depending on advisors to interpret it.\\

\subsection{Governance Cadence and Deviation Management}
\label{subsection: Governance Cadence and Deviation Management}

The governance architecture of the on-platform organisation operates at multiple frequencies, each calibrated to the nature of the events it is designed to manage.\\

At the operational level, continuous monitoring by the RTA detects deviations from project plans as they occur - triggering responses that are predefined and proportionate to the severity of the deviation. Expected deviations are managed through service level agreements calibrated to the estimated impact of each deviation type. Unexpected deviations are incorporated into the RTA as new pains competing for resources within the general prioritisation system - preventing incident management from operating as a parallel and ungovernable channel outside the governed transformation architecture.\\

At the tactical level, the AI Committee - composed of the CTO, CAIO, Chief Transformation Officer, and AI Champions - reviews the transformation portfolio on a bi-monthly basis, using the platform's self-drawn Three-Layer Company Model overview as the primary instrument. The review does not depend on external presentations or advisor interpretations: the organisation reads its own transformation data directly, adjusting RTA inputs where the review reveals new information that the algorithm has not yet incorporated.\\

At the strategic level, the board conducts quarterly reviews of annual KPIs and an annual review of the master transformation plan. The annual review operates with long-term strategic coherence as its governing principle, adjusted by the short-term data and accumulated learning that the continuous monitoring and tactical reviews have generated. Rigidity in the master plan is as dangerous a failure mode as its absence: the governance architecture must be capable of incorporating learnings without losing strategic coherence, and the RTA provides the mechanism through which this balance is maintained.\\

Deviations from the master plan are not treated as failures to be minimised at all costs. Governed deviations - deviations that arise within the established framework and are managed through the RTA rather than outside it - are a constitutive property of the federated transformation model. The equilibrium between centralisation and decentralisation that the Three-Ring Architecture instantiates necessarily generates deviations as local units adapt to their specific operational contexts. The governance architecture is designed to make these deviations visible, manageable, and productive rather than to eliminate them.\\

\subsection{Section Conclusions}
\label{subsection: Section 5 Conclusions}

The Three-Layer Company Model and the Ranked Transformation Agenda constitute the governance architecture that converts the diagnostic outputs of Section \ref{section: Extreme-efficient Nations} into a sequenced, resourced, and compounding transformation agenda. The Three-Layer Company Model provides the strategic classification framework - Tech Must, Right to Play, Right to Win - through which the board maintains a principled view of the transformation portfolio independent of departmental lobbying and advisory solution sets. The RTA provides the dynamic prioritisation algorithm through which that framework is applied to the organisation's specific context, incorporating quantitative impact assessment, calibrated heuristics, game theory, and synergy weighting into a continuously updated ranking that minimises Time-to-Production across the full project portfolio.\\

The compounding effect of the on-platform model - in which each project reduces the cost and increases the ROI of subsequent ones through natively inheritable architectural components - is the property that most clearly distinguishes the Three-Ring Architecture from the isolated project model it replaces. It is also the property that makes the governance architecture of this section analytically inseparable from the architectural foundation of Section \ref{section: The Three-Ring Architecture}: the compounding effect is a consequence of the shared Ring 2 infrastructure, and the RTA is the instrument through which that consequence is made explicit, governed, and maximised over time.\\

\section{Algorithmization Training}
\label{section: Algorithmization Training}

\subsection{The Structural Priority of Training}
\label{subsection: The Structural Priority of Training}

Enterprise AI transformation is not primarily a technological event. It is an epistemic one. The Three-Ring Architecture provides the infrastructure through which intelligence is governed and compounded. The RTA provides the prioritisation instrument through which transformation projects are sequenced. Neither produces transformation without the human layer capable of operating them with sufficient understanding to govern what they produce.\\

Training is therefore not a project ranked against other transformation projects within the RTA. It is the epistemic prerequisite of the RTA's own effective operation. Its structural priority is non-negotiable - not because it scores highest on the impact-effort matrix, but because it is the condition under which the matrix can be applied with the judgement the architecture requires.\\

This structural priority does not imply strict temporal precedence. The specific content, depth, and sequencing of training across profiles is governed by the RTA, calibrated to the organisation's diagnostic output and transformation trajectory. The priority of training as a category is structural. The prioritisation of specific training decisions within that category is governed and continuously adjusted.\\

\subsection{Algorithmization as the Conceptual Foundation}
\label{subsection: Algorithmization as the Conceptual Foundation}

The training the Three-Ring Architecture requires is not generic AI literacy. Generic AI literacy - familiarity with large language models, basic prompt engineering, awareness of AI use cases - is a right-to-play capability. It is necessary but not sufficient for the governance of a federated algorithmic architecture.\\

The required conceptual foundation is Algorithmization: the discipline that integrates business expertise, machine learning methods, and technology architecture into a unified framework for the design, deployment, and governance of algorithmic intelligence at enterprise scale \cite{Data MAPs}. Its distinguishing property is integrative: it trains all profiles in the intersection of the three domains - the sweet spot at which optimal algorithmic architecture requires simultaneous depth in business logic, statistical reasoning, and technical implementation \cite{Data MAPs}.\\

The deepest transformation Algorithmization training produces is cultural rather than technical: the replacement of a tool-adoption culture with a protocol-design culture. An organisation whose employees understand that technology is the instantiation of their own protocols - that the frontend they design and the if-then logic they contribute are the intelligence layer of the architecture - is structurally ready for the Three-Ring Architecture. The transition to this culture is the most consequential and most underestimated dimension of enterprise AI transformation.\\

A fourth strategic function of Algorithmization training is the development of internal judgement sufficient to evaluate, challenge, and govern external providers. The architectural errors documented in Subsection \ref{subsection: Section 3 Conclusions} are not only a consequence of vendor incentive structures. They are equally a consequence of the absence of internal judgement capable of recognising and resisting them. A Ring 3 vendor proposing direct Ring 1 integration succeeds not only because it has a commercial interest in doing so, but because the organisation lacks the architectural literacy to identify the governance gap that direct integration produces. Algorithmization training closes that gap - not by making the organisation hostile to external providers, but by making it capable of engaging them on architecturally informed terms: evaluating proposals, challenging assumptions, identifying architectural incompleteness, and governing the integration of external capability into its own architecture on its own terms. The transformation is, in this precise sense, finally controlled beyond providers. Internal judgement - not vendor narrative - becomes the governing instrument of the transformation agenda. This is the condition under which the RTA can function as designed: a prioritisation algorithm driven by the organisation's own empirically grounded assessment of its needs, resistant to the commercial distortions that advisor-mediated and vendor-mediated diagnostics systematically introduce.\\

\subsection{Training Architecture Across Profiles}
\label{subsection: Training Architecture Across Profiles}

\textbf{The governance layer} - AI Champions and executive leadership - receives mandatory preparation. AI Champions must understand the platform sufficiently to audit external technical teams, maintain coherence between developments and defined protocols, and translate between operational needs and architectural capabilities. Executive leadership must govern the transformation portfolio - the Three-Layer Company Model, the RTA, the right-to-play versus right-to-win distinction - without dependence on external interpretation.\\

\textbf{The workforce layer} receives voluntary preparation, enabled by a platform architecture whose marginal cost of participant onboarding approaches zero. Voluntary participation is strategically motivated: training experienced as imposed generates resistance; training experienced as accessible capability development generates genuine adhesion. Forum structures associated with each programme surface talent with transformational leadership capacity that formal identification processes systematically miss.\\

\textbf{The noise management layer} addresses a structurally underestimated risk: the continuous stream of commercially motivated narratives from Ring 3 vendors, media coverage, and advisory relationships that distort internal judgement and generate reactive decision-making disconnected from the organisation's actual architectural needs. Periodic ecosystem interpretation updates - distinguishing empirically grounded signal from commercially motivated noise - are an organisational hygiene measure whose value grows in direct proportion to the velocity of the external narrative environment. The development of this filtering capacity at the organisational level is not separable from the Algorithmization training programme: it is the applied expression of the architectural literacy that training produces, directed outward toward the provider ecosystem rather than inward toward the platform.\\

\subsection{Section Conclusions}
\label{subsection: Section 6 Conclusions}

Training occupies a structurally prior position in the transformation architecture because it is the epistemic prerequisite of all other components. Without the conceptual foundation of Algorithmization, the RTA cannot be operated with the judgement the architecture requires, AI Champions cannot perform the governance functions on which federated transformation depends, the cultural transition from tool-adoption to protocol-design cannot be completed, and the organisation remains vulnerable to the commercial distortions of the provider ecosystem it must govern.\\

Training is not a phase that precedes transformation. It is a continuous function that governs the human layer of the architecture throughout the full transformation trajectory - calibrated by the RTA, evolving with the transformation, and compounding in value as the organisation's architectural maturity and its capacity for independent judgement develop in parallel.\\

\section{Two Compatible Technology Paths - Modular or Integral}
\label{section: Two Compatible Technology Paths - Modular or Integral}

\subsection{The Technology Choice and Its Governing Logic}
\label{subsection: The Technology Choice and Its Governing Logic}

The Three-Ring Architecture does not prescribe a single technology deployment path. Organisations entering the transformation process differ along dimensions that are analytically significant for how the architecture is deployed: digital maturity, budget constraints, regulatory context, tolerance for operational disruption, and the degree to which the governance layer established through training has developed sufficient depth to support full integral deployment.\\

The two technology paths - modular and integral - are compatible rather than sequential. An organisation may pursue modular deployments in departments where right-to-play capability is the correct strategic ambition, while simultaneously pursuing integral deployment in departments where right-to-win development is the priority. The Transformation Bubbles methodology established in prior work makes this parallelism explicit: different departments may occupy different positions on the modular-to-integral spectrum simultaneously, all governed by the same RTA. The choice between paths is not made once at the outset of transformation - it is made continuously, department by department, project by project, as the RTA identifies the deployment mode that maximises value given the organisation's current architectural maturity and resource constraints.\\

The governing principle of the technology choice is the right-to-play versus right-to-win distinction established in Subsection \ref{subsection: The Three-Layer Company Model}. Modular deployment is the correct instrument for right-to-play capability acquisition. Integral deployment is the correct instrument for right-to-win development. The RTA governs the allocation of resources between the two paths by reference to this distinction, ensuring that right-to-play investment does not consume the resources required for right-to-win development, and that right-to-win ambition is not pursued without the architectural foundations that integral deployment requires.\\

\subsection{Modular - Standard Technology, Right-to-Play}
\label{subsection: Modular - Standard Technology, Right-to-Play}

The modular path operates at the standard technology layer: off-the-shelf applications, widely known methodologies, and legacy-compatible deployments that require no proprietary architectural commitment. Its defining properties are summarised in Table \ref{tab:Table 1}.

\begin{table}[h]
\centering
\begin{tabular}{|c|c|}
\hline
\textbf{Dimension} & \textbf{Modular} \\
\hline
Technology type & Standard · off-the-shelf \\
\hline
Architecture & Legay · remains as-is \\
\hline
Knowledge base & Widely known \\
\hline
Competitive position & Right-to-play \\
\hline
Organisational outcome & Capability acquired \\
\hline
\end{tabular}
\caption{Modular technology stacking leads to a standard state across dimensions}
\label{tab:Table 1}
\end{table}

The modular path brings the organisation to the competitive baseline of its sector without generating differential advantage. Existing systems remain as-is; the organisation acquires capability without transforming the infrastructure through which that capability is governed. This is not a limitation of implementation quality. It is a structural property of the solution type: standard technology, deployed without a governing platform layer, cannot compound. Each module remains bounded by its own scope, and the organisation's capability accumulates additively rather than multiplicatively.\\

The strategic value of the modular path is precisely its boundedness. It generates quick wins, establishes organisational credibility, and builds architectural familiarity. It is the correct choice for functions where right-to-play is the appropriate strategic ambition - and it is a legitimate long-term position for those functions, not merely a transitional one. The error is not deploying modular solutions where they are appropriate. The error is deploying them where integral deployment is required for competitive differentiation, and treating the right-to-play baseline as a right-to-win achievement.\\

The structural limitation of the modular path reproduces, at smaller scale, the architectural pattern identified in prior work as the source of isolated digital impacts without synergy exploitation \cite{Data MAPs}: each module bears its full development cost independently, no module benefits from the architectural foundations established by prior modules, and the organisation's transformation portfolio accumulates as a collection of disconnected investments rather than a compounding architecture.\\

\subsection{Integral - Advanced Platform, Right-to-Win}
\label{subsection:  Integral - Advanced Platform, Right-to-Win}

The integral path operates at the advanced technology layer: a custom platform architecture, proprietary by design, that transforms the organisation's relationship to its own intelligence rather than merely extending its existing capability. Its defining properties are summarised in Table \ref{tab:Table 2}.

\begin{table}[h]
\centering
\begin{tabular}{|c|c|}
\hline
\textbf{Dimension} & \textbf{Integral} \\
\hline
Technology type & Advanced · custom \\
\hline
Architecture & Platform · sovereign \\
\hline
Knowledge base & Proprietary \\
\hline
Competitive position & Right-to-win \\
\hline
Organisational outcome & Becomes technology \\
\hline
\end{tabular}
\caption{Integral technology creation leads to the state-of-the-art}
\label{tab:Table 2}
\end{table}

The organisation does not adopt technology. It becomes technology: its protocols, heuristics, and operational logic are algorithmically instantiated in a platform that is architecturally sovereign and continuously compounding. The distinction from the modular path is not one of degree but of kind: integral deployment produces a fundamentally different organisational asset - one that grows in value with every project deployed on it, that protects the intellectual property embedded in it by architectural design, and that generates differential advantage that is not replicable through standard technology stack composition.\\

The integral path is instantiated through three interdependent components - the Triplet.\\

\textbf{The first component} is consultancy on new workflows unlocked by technology. Domain professionals and AI Champions define how their area wants to work going forward - the if-then protocols that encode operational logic, regulatory knowledge, and strategic judgment - and the platform is configured to instantiate those protocols algorithmically. The consultancy function does not prescribe workflows. It provides the technical and transformational expertise through which the organisation's own domain knowledge is translated into platform-deployable protocols. The business leads; the platform follows.\\

\textbf{The second component} is the creation of the custom interface. The frontend is defined by the organisation - how it wants to work, how it wants to visualise its intelligence, how it wants to interact with the platform's capabilities. The backend architecture adapts to that design. This inversion - frontend defined by the business, backend adapted by the platform - is the architectural expression of the business-led principle that governs the Three-Ring Architecture throughout. It is also the mechanism through which the platform generates the organisational ownership that is the most effective structural reducer of resistance to change.\\

\textbf{The third component} is integration with the legacy architecture via Extended Production Architectures. EPAs are the technical mechanism through which Ring 2 connects to Ring 1 without disrupting it: bespoke development environments that coexist with legacy systems, maximising speed-to-production while preserving the organisation's full control over upgrades, software onboarding, and the removal of any component at any time. Ring 2 agents interact with Ring 1 on behalf of users - the new interface operates in Ring 2, the legacy system continues to operate in Ring 1 - until the legacy function has been sufficiently absorbed that direct Ring 1 interaction is no longer required for that function.\\

\subsection{The Federated Cost Model}
\label{subsection:The Federated Cost Model}

The economic structure of the integral path is designed to reflect the architectural logic of the Three-Ring Architecture. It is governed by the federated cost model: a payment architecture in which individual AI projects deployed on the platform fund the platform infrastructure itself.\\

A minimum upfront commitment establishes the architectural foundation of Ring 2. The remainder of the platform investment is unlocked progressively as agreed impact KPIs - defined at the outset of deployment - are reached in production. The organisation commits to the platform as the platform proves its impact in its specific operational context, rather than committing to capability that has not yet been validated. This systematic protocol to release technology funding until the software fees are completed generates a coordination mechanism with the C-suite and board that is both controlled - the maximum cost is bounded by the entire fee - and timely: value-triggered releases do not require board meetings to authorise incremental commitments that the KPI achievement has already justified.\\

The strategic consequence is analytically precise. The first projects deployed on the platform generate the value that funds the platform's continued deployment, which in turn reduces the cost and increases the ROI of subsequent projects. ROI is neutral in the first projects and structurally positive thereafter: each subsequent project inherits the architectural foundations established by prior ones without bearing their full development cost. The Ls fund the M. The larger the M, the lower the marginal cost of each subsequent L and the higher its ROI. Transformation accelerates as the platform matures - not because investment increases, but because the architectural foundation makes each subsequent investment more productive than the last.\\

This economic structure is the financial expression of the compounding effect established in Subsection \ref{subsection: Synergies and the Compounding Effect}. It is also the mechanism through which the right-to-win ambition of the integral path is made financially accessible to organisations that cannot sustain the full platform investment as an upfront cost: value generation and platform investment are coupled rather than decoupled, ensuring that the organisation's financial commitment scales with demonstrated architectural value rather than preceding it.\\

The model admits two further extensions that become relevant as the platform fee grows in proportion to the organisation's P\&L. The first concerns the payment instrument. When the platform fee represents a sufficiently high share of the organisation's P\&L, a pure cash payment model concentrates financial burden at the moment of lowest organisational capacity to bear it - before the compounding value of the platform has fully materialised. In this regime, a cash-and-stock split is the analytically appropriate instrument: the provider accepts a portion of its fee in the organisation's equity, distributing the financial burden across the transformation horizon and aligning the provider's return with the long-term value the transformation generates. The greater the stock component, the stronger the incentive alignment: the provider's financial outcome becomes a direct function of the quality and depth of the transformation it enables, rather than a function of contractual KPI achievement alone. This alignment is architecturally consistent with the Three-Ring Architecture's logic — the platform compounds the organisation's value, and the provider participates in that compounding.\\

The second extension concerns the direction of investment. When the organisation's dependence on the provider is sufficiently structural - when the provider's frontier capability, as established in Subsection \ref{subsection: The Centre of Excellence}, is a necessary input to the organisation's right-to-win competitive position — the organisation should be permitted to invest in the provider. This converts the relationship from a vendor engagement into a strategic partnership with mutual compounding incentives: the organisation's investment funds the frontier innovation that the organisation absorbs through the centre of excellence relationship; the provider's equity stake in the organisation aligns its incentives with the organisation's long-term success. The two extensions are therefore not merely financial arrangements. They are the economic expression of the role architecture developed in Section \ref{section: Roles for Absorbing AI in the On-Platform Organisation}: the centre of excellence and the on-platform organisation are not transacting parties but compounding partners, and the capital structure of their relationship should reflect that property.\\

\subsection{The Governed Path Between Paths}
\label{subsection:The Governed Path Between Paths}

The modular and integral paths are not mutually exclusive choices made once at the outset of transformation. They are compatible positions on a governed transformation architecture, and the RTA determines their allocation across the organisation's function portfolio continuously as the transformation evolves.\\

Most organisations will operate both paths simultaneously: modular deployments in functions where right-to-play is the appropriate ambition, integral deployment in functions where right-to-win development is the priority. The RTA governs the boundary between the two - identifying, for each function and each project, which path generates more value given the organisation's current architectural maturity, resource constraints, and competitive position. This is not a binary choice but a continuous allocation problem, and the RTA is the instrument through which it is solved with analytical rigour rather than organisational politics.\\

The transition from modular to integral - where the RTA identifies it as value-generating - is not a discontinuous architectural event. Modular deployments contribute to the organisation's architectural familiarity and governance capacity, preparing the conditions for integral deployment without requiring it. The governance structures established in the modular phase - AI Champions, RTA protocols, Three-Layer Company Model classifications - migrate onto the platform without disruption. The organisation extends its transformation rather than restarting it.\\

The destination is invariant across both paths: the on-platform organisation, in which all working protocols, intelligence processes, and governance mechanisms operate within the federated architecture of the Three-Ring Architecture, governed by Ring 2, driven by the business, and compounding over time. The path varies. The destination does not.\\

\subsection{Section Conclusions}
\label{subsection:Section 7 Conclusions}

The two technology paths - modular and integral - are compatible instruments calibrated to the right-to-play versus right-to-win distinction that governs the transformation agenda. Modular deployment acquires standard capability efficiently, maintaining legacy systems as-is and bringing the organisation to the competitive baseline of its sector. Integral deployment transforms the organisation into a proprietary, compounding algorithmic architecture - sovereign, auditable, and continuously generating differential advantage that standard technology stack composition cannot replicate.\\

The choice between paths is not architectural preference. It is a governed prioritisation decision, made continuously by the RTA across the organisation's function portfolio, calibrated to diagnostic outputs, resource constraints, and the competitive position that each function requires. The federated cost model ensures that the financial structure of the integral path reflects its architectural logic: value generation and platform investment are coupled, and the compounding effect of the on-platform architecture makes each subsequent investment more productive than the last.\\

Both paths are governed by the same epistemic foundation - Algorithmization training - and oriented toward the same destination: the on-platform organisation in which intelligence compounds across departments and time, controlled beyond providers, and driven by the business rather than by the technology.\\

\section{Roles for Absorbing AI in the On-Platform Organisation}
\label{section: Roles for Absorbing AI in the On-Platform Organisation}

\subsection{Absorption as a Permanent Organisational Function}
\label{subsection: Absorption as a Permanent Organisational Function}

The deployment of the Three-Ring Architecture does not complete the transformation of the organisation into an on-platform entity. It provides the infrastructure through which transformation becomes continuous. The on-platform organisation is not a state that is reached and maintained - it is a state that must be actively reproduced through the permanent absorption of frontier AI capability into operational reality. That absorption requires a structured set of roles whose collective function is not episodic but continuous: not the management of a transformation programme but the institutionalisation of transformation as an organisational discipline.\\

This distinction has direct consequences for how the roles are designed. Roles structured around a transformation programme are transitional by definition - they are resourced for a phase and wound down when the phase concludes. Roles structured around the permanent absorption of frontier innovation are constitutive: they are as permanent as the organisation's need to remain at the frontier of its own intelligence layer. The question is not when the transformation will be complete. It is how the organisation structures itself to ensure that the latency between frontier innovation and operational impact remains minimal, continuously and at scale.\\

Four role categories govern this absorption process. Their order reflects the direction of innovation flow: from the frontier toward operational reality, through the layers of translation and execution that make frontier innovation productive within the organisation's specific context.\\

\subsection{The Centre of Excellence}
\label{subsection: The Centre of Excellence}

The centre of excellence is the frontier innovation layer of the on-platform organisation. Its function is to shift the knowledge frontier rather than follow it - a distinction that is analytically critical and operationally non-trivial. An organisation whose intelligence layer follows the frontier acquires right-to-play capability: it reaches the competitive baseline of its sector without generating differential advantage. An organisation whose intelligence layer shifts the frontier - contributing to the advancement of the frontier rather than merely adopting its outputs - acquires right-to-win capability: proprietary innovations that are not yet available to competitors and that are, by construction, not replicable through standard technology stack composition.\\

The scarcity of genuine centres of excellence is a structural property of the knowledge economy rather than a contingent feature of the current moment. The complexity of operating at the frontier of algorithmic knowledge - integrating business expertise, machine learning methods, and technology architecture at the depth required to produce innovations rather than applications - is extreme. There will be very few organisations on earth capable of performing this function at the level the Three-Ring Architecture requires. Side-of-excellence approximations do not constitute a viable substitute: they introduce noise rather than signal, consuming the organisation's absorption capacity without advancing its frontier position. The distinction between a genuine centre of excellence and a decorative one is not a matter of degree. It is a structural difference with direct consequences for whether the organisation's intelligence layer advances or stagnates.\\

Centres of excellence may be internal - embedded within the organisation's own research and development function - but are more commonly external partners: research institutions, deep-tech companies, or academic collaborations whose frontier position is independently maintained and whose relationship with the on-platform organisation is governed by a structural partnership rather than a conventional advisory engagement. Critically, centres of excellence impact the organisation through multiple vehicles: companies that are acquired, joint ventures, licensed innovations, or federated research programmes. The relationship is not consultative. It is generative - the centre of excellence produces the innovations that the organisation absorbs, and the organisation's operational reality produces the problems that the centre of excellence addresses.\\

\subsection{The Catalyser: Transformation and Evolution Department}
\label{subsection: The Catalyser: Transformation and Evolution Department}

The catalyser is the translation layer between the centre of excellence and the business. Its governing function is the minimisation of latency between frontier innovation and operational impact: the elapsed time between the moment a frontier innovation becomes available and the moment it is productively deployed within the organisation's operational architecture. This latency is the primary determinant of whether the organisation's frontier relationship generates competitive advantage or merely generates awareness of innovations that competitors are absorbing at the same speed.\\

The catalyser performs this function through three interdependent activities. The first is cultural bootstrapping: the translation of frontier innovations into a form that the organisation's culture and its individuals can absorb. Frontier innovations that are technically correct but culturally illegible to the organisation generate resistance rather than adoption - the latency between innovation and impact is as much a cultural variable as a technical one. The catalyser governs this cultural dimension explicitly, calibrating the pace and form of innovation absorption to the organisation's actual capacity rather than its theoretical readiness.\\

The second activity is protocol development: the translation of frontier innovations into the operational protocols, workflow redesigns, and heuristic structures that make them deployable within the Ring 2 architecture. This is not a passive translation function. It requires the catalyser to understand both the frontier innovation and the operational context with sufficient depth to identify the specific protocol changes that the innovation makes possible - and to design those changes in a form that the business can own and the data scientists can implement.\\

The third activity is self-improvement: the catalyser is itself subject to the continuous efficiency discipline it applies to the rest of the organisation. It chases better protocols, more Algorithmization, better training methodologies, and more effective cultural bootstrapping mechanisms - treating its own operational logic as a subject of the same algorithmic discipline it applies elsewhere. A catalyser that does not improve its own protocols cannot credibly govern the improvement of others.\\

The AI Champion population sits organisationally within the catalyser - or reports to it in addition to reporting to the business. This double reporting structure reflects the AI Champion's dual accountability precisely: to the business for operational governance and audit coherence, and to the catalyser for transformation coherence and frontier absorption. The AI Champion is the living interface between the catalyser's translation function and the business's operational reality - the role through which the latency minimisation discipline of the catalyser is instantiated at the departmental level.\\

\subsection{The Business}
\label{subsection: The Business}

The business - the domain expert layer of the on-platform organisation - is both the destination of frontier innovation and its operational governor. Domain experts define the protocols and heuristics that make frontier innovations operationally meaningful: without the business's if-then logic, frontier innovations remain technically capable but operationally inert. The business is not a passive recipient of innovations that the centre of excellence and the catalyser deliver. It is the active agent that determines which innovations are worth absorbing, how they should be instantiated in operational protocols, and what competitive advantage they can generate within the organisation's specific context.\\

This active role requires the business to maintain a continuous awareness of frontier shifts - not at the technical depth of the centre of excellence, but at the strategic depth required to identify which shifts have operational implications for the organisation's own intelligence layer. This awareness is developed through the relationship with the catalyser: the catalyser translates frontier shifts into strategically legible form; the business evaluates their operational implications and drives the absorption agenda through the RTA.\\

The business also governs the absorption process through the Three-Layer Company Model: classifying each innovation absorption project as Tech Must, Right to Play, or Right to Win, and allocating resources accordingly. The discipline of this classification - resisting the tendency to treat all frontier innovations as right-to-win opportunities, and identifying which innovations merely advance the sector baseline - is the business's primary contribution to the efficiency of the absorption process.\\

\subsection{Data Scientists and Data Engineers}
\label{subsection: Data Scientists and Data Engineers}

Data scientists and data engineers constitute the execution layer of the absorption process: the population that translates the combined outputs of the centre of excellence, the catalyser, and the business into production-grade algorithmic implementations deployable within the Ring 2 architecture.\\

Their role is not frontier innovation. They are not expected to shift the knowledge frontier - that is the centre of excellence's function. Their role is to take frontier innovations and operational heuristics and build the algorithmic implementations that make both productive within the organisation's specific operational context. This execution function is what allows the business to take genuine ownership of its intelligence layer beyond the centre of excellence: data scientists and data engineers embed frontier capability into the organisation's own architecture, making it proprietary, compounding, and independent of continued centre of excellence involvement for each subsequent implementation decision.\\

The generalist character of data science methods - transferable across domains, applicable to diverse operational contexts with relatively modest domain immersion - is the structural property that makes this execution layer scalable. Unlike the certified integrator ecosystems of prior enterprise platforms, the data science and data engineering talent pool is globally abundant, growing independently of any certification programme, and not controllable by any incumbent. The limiting factor on the execution layer's productivity is not the supply of technical capability. It is the quality of the operational heuristics and frontier innovations that the business and the centre of excellence supply - and the effectiveness of the catalyser in minimising the latency between those inputs and the execution layer's implementation of them.\\

\subsection{Federation as the Enabling Condition}
\label{subsection: Federation as the Enabling Condition}

The role architecture developed in this section - centre of excellence, catalyser, business, data scientists and data engineers - operates sustainably only under the governance conditions that the Three-Ring Architecture's federation layer provides. Federation is not a background property of the architecture. It is the enabling condition without which the role architecture cannot function at the scale, security, and cost structure that the on-platform organisation requires.\\

The federation layer provides five properties that are directly relevant to the sustainability of the absorption process. The first is operational robustness: the federated architecture ensures that the absorption process continues to function under conditions of disruption - technical failures, organisational changes, or external shocks - without systemic cascading consequences. The second is cybersecurity: IP protection is enforced at every node of the architecture by design, ensuring that frontier innovations absorbed into the organisation's intelligence layer are protected from external exposure without requiring post-hoc security additions. The third is sovereignty: the organisation retains full control over its own intelligence layer - its data, its models, its protocols - without dependence on any external vendor's infrastructure as a condition of operation.\\

The fourth property concerns the talent rotation dynamic that is a structural feature of the extreme-efficient organisations that on-platform architectures produce. Contemporary workforce dynamics - characterised in particular by high voluntary turnover rates among Generation Z technical profiles, low organisational loyalty, and abbreviated tenure expectations - create a systemic risk that is fractal in nature: every department within any organisation of any size is exposed to the same structural vulnerability at its own scale \cite{Embrace_rotation}. In extreme-efficient organisations, where the departure of even a single individual can constitute a systemic risk, this vulnerability is most acute. The federation layer addresses it structurally: knowledge resides in the platform rather than in the people. The controlled departure of technical profiles with compartmentalised access to the platform reinforces rather than weakens the organisation's proprietary intelligence asset - because no individual accumulates a complete view of the system, and the intelligence they helped build remains in the architecture after their departure. Talent rotation, managed through the federated architecture, is transformed from a systemic risk into an antifragility mechanism \cite{Embrace_rotation}.\\

The fifth property is cost efficiency: the federated architecture scales at the organisation's pace rather than against it. The marginal cost of each subsequent project decreases as the platform matures; the compounding effect established in Subsection \ref{subsection: Synergies and the Compounding Effect} ensures that the absorption process becomes progressively cheaper and faster as the intelligence layer grows. The role architecture does not become more expensive as the organisation's frontier ambition increases. It becomes more productive - because each new centre of excellence innovation, each new catalyser protocol, each new business heuristic, and each new data science implementation builds on an architectural foundation that prior contributions have already validated and compounded.\\

\subsection{The BusinessSection Conclusions}
\label{subsection: Section 8 Conclusions}

The absorption of AI into the on-platform organisation is a permanent organisational function that requires a structured role architecture calibrated to the direction of innovation flow: from the frontier toward operational reality, through the layers of translation and execution that make frontier innovation productive. The centre of excellence shifts the frontier; the catalyser minimises the latency between frontier innovation and operational impact; the business governs the absorption agenda and defines the operational heuristics that make innovations productive; data scientists and data engineers execute the implementations that embed frontier capability into the organisation's proprietary intelligence layer.\\

The AI Champion sits at the intersection of the catalyser and the business - double reporting, dual accountability, the living interface between frontier absorption and operational reality. The quality of this role architecture is the primary determinant of the organisation's capacity to remain at the frontier of its own intelligence layer continuously rather than episodically.\\

The federation layer is the enabling condition of the entire architecture: providing the robustness, cybersecurity, IP protection, talent rotation management, and cost efficiency that allow the role architecture to operate sustainably at scale. Without federation, the absorption process is vulnerable to the systemic risks that talent rotation, cybersecurity exposure, and vendor dependency introduce. With federation, those risks are governed by architectural design - and the organisation's intelligence layer compounds continuously, driven by the business, protected by the platform, and advancing toward the frontier that the centre of excellence defines.\\

\section{Conclusions and Future Work}
\label{section: Conclusions and Future Work}

\subsection{Conclusions}
\label{subsection: Conclusions}

This paper has argued that the governing failure of enterprise AI transformation is architectural rather than capability-related, and that the correct architectural response is the Three-Ring Architecture - a federated infrastructure that constitutes, in the technically exact sense of the term, the operating system of the on-platform organisation.\\

The argument has been developed across eight sections whose contributions can be stated precisely.\\

The OS analogy is technically exact. Ring 2 performs, at the enterprise level, the same four functions - resource abstraction, process coordination, permission enforcement, and platform provision - that a computing OS performs at the device level. Prior enterprise software did not constitute an OS in this sense: it was application software that required specialist integrators external to the business and generated switching-cost moats rather than compounding intelligence. Ring 2 constitutes the missing infrastructure layer - the substrate on which the intelligence of the enterprise is governed, coordinated, and compounded.\\

The theoretical necessity of Ring 2 is established on two independent grounds. The butterfly problem - the sensitive dependence on initial conditions that characterises any deterministically complex system - generates a class of governance requirement that cannot be met at the application layer. The dice problem - non-determinism at the node level introduced by LLM-based agents - generates a categorically distinct and more severe governance requirement for which no application-level remedy exists. Ring 2 is not a useful addition to an enterprise AI architecture that includes Ring 3. It is a necessary condition of that architecture's operational safety. The strength of this necessity scales with the capability of Ring 3: more capable non-deterministic actors generate more consequential deviations and therefore require more robust governance.\\

The current market configuration reproduces, at software speed, the identical architectural error that defined the first wave of enterprise AI deployment: decentralised intelligence without a federation layer. Ring 3 vendors connecting directly to Ring 1 - bypassing Ring 2 - are not providing a path to production-grade transformation. They are providing extraordinary capability without the governing infrastructure that makes that capability safe, auditable, and compounding at enterprise scale. This pattern is not coincidental: transformation is becoming a byproduct of LLM capability rather than the object of independent architectural design, driven by commercial positioning rather than by the systematic study of what production-grade transformation requires \cite{Data MAPs}. Vendor neutrality is an architectural prerequisite of Ring 2, and neutrality is structurally incompatible with being a Ring 3 vendor. The governing layer cannot be supplied by the intelligence layer it governs.\\

The diagnostic foundation of the EEN framework addresses the epistemic limitations of self-assessment and advisor-mediated diagnostics by providing a cross-sector empirical benchmark against which any organisation can locate its actual transformation position. Its sector benchmark distinguishes right-to-play improvements from right-to-win opportunities - the most strategically significant output of the diagnostic process. At the national level, the same aggregation logic that calibrates individual transformation agendas constitutes the empirical substrate of evidence-based economic policy: a bottom-up signal of real organisational needs that inverts the conventional top-down policy design logic, grounding national economic strategy in the operationally defined needs of the enterprise population it is designed to serve.\\

The governance architecture of the Three-Layer Company Model and the Ranked Transformation Agenda converts diagnostic outputs into a sequenced, resourced, and compounding transformation agenda. The RTA's synergy-weighted prioritisation - in which each project reduces the cost and increases the ROI of subsequent ones through natively inheritable architectural components - is the mechanism through which the on-platform organisation's intelligence compounds rather than accumulates. The fractal governance principle ensures that the transformation is governed on the platform from the first day of deployment: the architecture governs its own construction.\\

Training occupies a structurally prior position in the transformation architecture because it is the epistemic prerequisite of all other components. Algorithmization training produces the conceptual foundation - integrating business expertise, machine learning methods, and technology architecture - without which the RTA cannot be operated with the judgement the architecture requires, AI Champions cannot perform the governance functions on which federated transformation depends, and the organisation remains vulnerable to the commercial distortions of the provider ecosystem. Critically, Algorithmization training develops the internal judgement required to evaluate, challenge, and govern external providers - placing the transformation finally under the control of the organisation rather than its advisors.\\

The two technology paths - modular and integral - are compatible instruments calibrated to the right-to-play versus right-to-win distinction. Modular deployment brings the organisation to the competitive baseline of its sector efficiently, maintaining legacy systems as-is. Integral deployment - instantiated through the Triplet of workflow consultancy, custom interface creation, and EPA legacy integration - transforms the organisation into a proprietary, compounding algorithmic architecture that is sovereign, auditable, and continuously generating differential advantage. The federated cost model ensures that the financial structure of the integral path reflects its architectural logic: individual projects fund the platform infrastructure, and the platform compounds the value of each subsequent project. Where the platform fee represents a high proportion of the organisation's P\&L, a cash-and-stock split is the analytically appropriate instrument: the provider's return becomes a function of the transformation's long-term value generation rather than contractual KPI achievement alone. Where the organisation's dependence on the provider's frontier capability is structural, the organisation should be permitted to invest in the provider - converting a vendor relationship into a strategic partnership with mutual compounding incentives whose capital structure reflects the depth of the architectural interdependence.\\

The absorption of AI into the on-platform organisation is a permanent organisational function, not a transitional phase. It requires a structured role architecture calibrated to the direction of innovation flow. The centre of excellence shifts the knowledge frontier - there will be very few of these on earth, and side-of-excellence approximations introduce noise rather than signal. The catalyser minimises the latency between frontier innovation and operational impact, bootstrapping the relationship between frontier knowledge and operational deployment calibrated to the organisation's culture; AI Champions sit within the catalyser or report to it in addition to the business, constituting the living interface between frontier absorption and operational reality. The business governs the absorption agenda and defines the operational heuristics that make innovations productive. Data scientists and data engineers execute the implementations that embed frontier capability into the organisation's proprietary intelligence layer, enabling the business to take genuine ownership of its intelligence beyond the centre of excellence. This role architecture operates sustainably only under the governance conditions that federation provides: robustness, cybersecurity, IP protection at every node, and the structural management of talent rotation as an antifragility mechanism rather than an organisational risk. Contemporary workforce dynamics - characterised by high voluntary turnover rates among Generation Z technical profiles - constitute a systemic risk that is fractal in nature: every department within any organisation of any size is exposed to the same structural vulnerability at its own scale \cite{Embrace_rotation}. The federated architecture addresses this structurally: knowledge resides in the platform rather than in the people, and the controlled departure of technical profiles with compartmentalised access reinforces rather than weakens the organisation's proprietary intelligence asset.\\

The market consequence of this framing is a structural repositioning of the addressable opportunity. The Three-Ring Architecture competes not for the software budget but for the labour budget - the cost of the intelligence work that humans and outsourced service providers currently perform across every department, every company, every industry. The OS's addressable market is every algorithmically-driven organisation on earth, which within a decade will be every organisation that survives. Every improvement in Ring 3 capability expands Ring 2's necessity rather than contracting it. The governing layer always benefits from more powerful - and more consequential - applications running on top of it.\\

We conclude with the observation that frames the entire analysis. The algorithmic enterprise is not a future state. It is the present trajectory of every organisation that intends to remain competitive. The question of whether enterprises will need an OS for their intelligence is already settled by the structural properties of the systems they are deploying and the regulatory environments they are operating in. The question that remains open is which OS they will run - and whether they will reach that destination through a governed architectural path or through the accumulation of isolated Ring 3 deployments that reproduce, at greater speed and cost, the failure pattern that has defined enterprise AI transformation since its inception.\\

\subsection{Future Work}
\label{subsection: Future Work}

The research agenda opened by this paper is substantial. Four primary directions are identified.\\

The first is the formalisation of the ecosystem economics of the Three-Ring Architecture: the development of a microeconomic model for the compounding effects of the developer ecosystem, including the rate at which organisational intelligence accumulates, the conditions under which it transfers across organisations in federated architectures, and the valuation implications of intelligence accumulation as an intangible asset. This work connects naturally to ongoing research on the valuation of algorithmically transformed companies - including the formal demonstration that controlled technical talent rotation, managed through compartmentalised platform architecture, constitutes an IP protection mechanism and an antifragility asset rather than an organisational risk \cite{Embrace_rotation}.\\

The second is the empirical development of the EEN diagnostic framework as a longitudinal instrument, extended to include the systematic study of absorption latency: the rate at which organisations with different role architectures - with and without a genuine centre of excellence relationship, with and without a catalyser function - move from frontier innovation to operational impact. A longitudinal dataset tracking the same population of organisations across multiple diagnostic cycles would produce the first empirically grounded account of transformation velocity and its structural determinants. This dataset would constitute a uniquely valuable resource for both transformation planning and investment due diligence, and would provide the empirical foundation for the role architecture argument developed in Section \ref{section: Roles for Absorbing AI in the On-Platform Organisation}.\\

The third is the capital structure of deep-tech frontier partnerships. The cash-and-stock payment model and the mutual investment argument developed in Subsection \ref{subsection:The Federated Cost Model} open a research direction in the economics of frontier partnerships that has not been systematically studied. The conditions under which stock consideration is analytically appropriate, the optimal allocation between cash and equity across different P\&L weight scenarios, and the governance implications of mutual investment between on-platform organisations and their frontier partners constitute a research programme at the intersection of corporate finance, strategic management, portfolio management and the economics of platform architectures.\\

The fourth is the extension of the OS framing to the national level - building on prior work in \cite{EEN}. If Ring 2 is the OS of the algorithmic enterprise, the question of national competitiveness in the algorithmic era reduces to the question of which nations adopt the Three-Ring Architecture at scale across their enterprise ecosystems first. The EEN framework's capacity to aggregate bottom-up organisational diagnostic signals into national policy instruments is the empirical foundation on which this programme rests. The geopolitical implications - for trade, productivity, defence, and the distribution of economic value in the algorithmic era - constitute a research programme whose scope extends well beyond the boundaries of this paper, and whose development represents the most consequential direction for future work that this paper identifies.\\

%%%%%%%%%%% Bibliography with links%%%%%%%%%%%%%%%%%%%%%%%%%%

%%%%%%%%%%%%%%%%%%%%%%%%%%%%%%%%%%%%%

%%%%%%%%%%% Bibliography %%%%%%%%%%%%%%%%%%%%%%%%%%
%\begin{thebibliography}{100}
% \bibitem{PhD} Álvarez-Teleña S., Systematic Trading: Calibration Advances through Machine %Learning (Doctoral Thesis, University College London, 2014)
% \bibitem{Data MAPs} Álvarez-Teleña S., Díez-Fernández M., Data MAPs: On-Platform %Organisations (SSRN, 2022)
% \bibitem{DDP} Álvarez-Teleña S., Díez-Fernández M., Advances in Portfolio Management: %Dimension-Driven Portfolios (SSRN, 2023)
% \bibitem{Performance Attribution} Álvarez-Teleña S., Díez-Fernández M., Advances in %Portfolio Management: Performance Attribution by Design (SSRN, 2023)
% \bibitem{OTAN} Álvarez-Teleña S., Díez-Fernández M., Advances in Cognitive Warfare: %Augmented Machines upon Data MAPs towards a Fast and Accurate Turnaround (SSRN, 2023)
%\bibitem{Nature} Park, M., Leahey, E. \& Funk, R.J., Papers and patents are becoming less %disruptive over time (Nature 613, 138–144, 2023)

%\end{thebibliography}
%%%%%%%%%%%%%%%%%%%%%%%%%%%%%%%%%%%%%

%\section{References}
%\label{section: References}

%\begin{enumerate}[label={[\arabic*]}]
%\item{Álvarez-Teleña S., Díez-Fernández M., Data MAPs: On-Platform Organisations (SSRN, 2022)}
%\end{enumerate}

\end{document}